%% file: paper.tex
\documentclass[twocolumn,aps,prd,floatfix,preprintnumbers]{revtex4}
\usepackage{graphicx} 
\usepackage{dcolumn}
\usepackage{epsfig}
\usepackage{amsmath}
\usepackage{color}
\usepackage{verbatim}

\input babarsym.tex   

\def\etal {{\it et al.}}

\mathchardef\myhyphen="2D

\newcommand{\SLACPubNumber}   {2510101}
\newcommand{\allm}{\ensuremath{\kern 0.15em }}
\newcommand{\all}{\ensuremath{\kern 0.25em }}
\newcommand{\al}{\ensuremath{\kern 0.5em }}
\newcommand{\alpp}{\ensuremath{\kern 0.75em }}
\newcommand{\alp}{\ensuremath{\kern 1.0em }}
\newcommand{\alx}{\ensuremath{\kern 10.0em }}
\newcommand{\allmm}{\ensuremath{\kern -1.25em }}
\newcommand{\almm}{\ensuremath{\kern -0.75em }}
\newcommand{\aln}{\ensuremath{\kern -0.25em }}
\newcommand{\alm}{\ensuremath{\kern -0.50em }}
\newcommand{\almx}{\ensuremath{\kern -1.50em }}
\def\itgamma {\ensuremath{\it \Gamma}\xspace}
\def\itdelta {\ensuremath{\it \Delta}\xspace}

\def\chibj{\ensuremath{\chi_{bJ}(2P)}\xspace}
\def\chibzero {\ensuremath{\chi_{b0}(2P)}\xspace}
\def\chibone  {\ensuremath{\chi_{b1}(2P)}\xspace}
\def\chibtwo  {\ensuremath{\chi_{b2}(2P)}\xspace}
\def\chib {\ensuremath{\chi_{b}(2P)}\xspace}
\def\chibtot {\ensuremath{\chi_{b1,2}(2P)}\xspace}
\def\chibtotn {\ensuremath{\chi_{b0,1,2}(2P)}\xspace}
\def\calB         {{\ensuremath{\cal B}\xspace}}
\def\mrec {\ensuremath{m_{rec}(\gamma_s)}\xspace}
\def\calL         {{\ensuremath{\cal L}\xspace}}
\def\ee {\ensuremath{\ep\en}\xspace}
\def\evtgen     {\mbox{\textsc{EvtGen}}\xspace}
\begin{document}

\begin{flushleft}
  SLAC-PUB-\SLACPubNumber \\
\end{flushleft}
  
\title{
 \large \bf\boldmath Study of $\chi_{b1,2}(2P) \to \omega \Upsilon(1S)$ transitions in $\Upsilon(3S) \to \gamma \chi_{b1,2}(2P)$ decays at \babar
}

\input  authors_aug2025_frozen_no_institutes

\begin{abstract}
  \noindent
  
Results are presented on $\chi_{b1,2}(2P) \to \omega \Upsilon(1S)$ transitions from $\epem \to \Upsilon(3S) \to \gamma \chi_{b1,2}(2P)$ decays. The data were collected
with the \babar\ detector at the PEP-II asymmetric-energy $e^+e^-$ collider at SLAC. The integrated
luminosity of the data sample is 28.0 fb$^{-1}$, corresponding to $121.3 \times 10^6$ $\Upsilon(3S)$ decays.
Signals of $\chi_{b1,2}(2P)$ are observed over a negligible background. Improved precision measurements of branching fractions are obtained. First measurements of the $\chi_{b1,2}(2P)$ angular distributions are performed.
No evidence is found for the presence of a $\chibzero \to \omega \OneS$ decay mode.  
\end{abstract}

\maketitle

\section{Introduction}
\label{sec:intro}

The strongly bound $b \bar b$ meson system, bottomonium, exhibits a rich positronium-like structure that is a laboratory for verifying perturbative and nonperturbative
QCD calculations~\cite{Besson:1993mm,Eichten:2007qx,Brambilla:2010cs}. Potential models and lattice calculations provide good descriptions of the mass structure and radiative transitions below the open-flavor threshold.
Precision spectroscopy probes spin-dependent and relativistic effects. Quark-antiquark potential formulations
have been successful at describing the bottomonium system phenomenologically~\cite{Besson:1993mm,Eichten:2007qx,Brambilla:2010cs}.

In 2004, the CLEO Collaboration~\cite{CLEO:2003vio} reported the first observation of the transitions $\chi_{bJ}(2P)\to\omega\Upsilon(1S)$, produced in radiative decays of $(5.81\pm0.12)\times10^6~\Upsilon(3S)$ mesons. The \chib branching fractions of the $J=1$ and $J=2$ states were measured to be $(1.63^{+0.35+0.16}_{-0.31-0.15})\%$ and $(1.10^{+0.32+0.11}_{-0.28-0.10})\%,$ respectively.  
The implications of these results are discussed in Ref.~\cite{Voloshin:2003zz}.

The observation of the near-threshold transition of the $c \bar c$ state $\chi_{c1}(3872) \to \omega \jpsi$ by the Belle Collaboration~\cite{Belle:2004lle} is of particular interest.
Although it is a narrow state, $\Gamma(\chi_{c1}(3872))=1.19\pm0.21$ MeV~\cite{PDG}, 
that lies nearly 8 MeV below the nominal kinematic $\omega J/\psi$ threshold, the observed branching fraction is nearly as large as that of the discovery channel $\chi_{c1}(3872) \to \pip \pim J/\psi$.

A similar decay mode could exist for the $b \bar b$ state $\chibzero \to \omega \OneS$
given the predicted width of about 2.6 \mev\ for this resonance~\cite{Godfrey:2015dia}.
Given the low significance of the reconstructed signal, no search was performed by CLEO for a \chibzero contribution, expected to be strongly suppressed by the $\omega \OneS$ mass threshold.
More recently, $\chibj\to \omega \OneS$ decay modes have been confirmed by Belle~\cite{Belle:2024azd} with additional evidence for a \chibzero decay to this final state.
Precise measurements of branching fractions for radiative decays between bottomonium states have been reported, recently, by the \babar\ Collaboration~\cite{BaBar:2014och}.

\section{The \babar\ detector and dataset}
\label{sec:babar}

The results presented here are based on a data sample of 28.0 \invfb of integrated
  luminosity~\cite{lumy} collected at the \ThreeS resonance with the \babar\ detector at the
  PEP-II asymmetric-energy  $e^+e^-$  collider at SLAC.
The \babar\ detector is described in detail elsewhere~\cite{BABARNIM}.
The momenta of charged particles are measured
by means of a five-layer, double-sided microstrip detector
and a 40-layer drift chamber, both operating  in the 1.5~T magnetic 
field of a superconducting solenoid. 
Photons are measured and electrons are identified in a CsI(Tl) crystal
electromagnetic calorimeter (EMC). Charged-particle
identification is provided by the measurement of specific energy loss in
the tracking devices and by an internally reflecting, ring-imaging
Cherenkov detector. 
Muons and \KL\ mesons are detected in the
instrumented flux return  of the magnet.  Decays of unstable particles
  are described by \evtgen~\cite{Lange:2001uf}. 
Monte Carlo (MC) simulated events~\cite{geant}, with sample sizes 
around 20 times larger than the corresponding data samples, are
used to evaluate the signal efficiency.
Final-state radiation effects are described by PHOTOS~\cite{Barberio:1993qi}.

\section{Event reconstruction} 
\label{sec:reco}

The following decay chain is reconstructed
\begin{equation}
\begin{array}{ll}
  \ThreeS \to \gamma_s \chib & \\
  &\almx \to \omega \OneS,\\
  \end{array}
\label{eq:sig}
\end{equation}
where $\omega \to \pip \pim \piz$, $\OneS \to \ellp \ellm$, $\ell=\electron,\mu$, and $\gamma_s$ indicates
the ``signal'' $\gamma$ to distinguish it from background $\gamma$'s and those from the \piz\ decay.
Only events containing exactly four well-measured
tracks with transverse momentum greater than 0.1~\gevc\ 
and a total net
charge equal to zero are considered.
The four tracks 
are fitted to a common vertex, with the requirements that the fitted vertex be within the
$e^+ e^-$ interaction region and have a $\chi^2$ fit probability greater than 0.001.
Well-reconstructed $\gamma$'s in the EMC with an energy greater than 30 MeV  are reconstructed, retaining up to 20 candidates per event.
They are combined to form \piz\ candidates. Combinations having an invariant mass in the
(0.115-0.155) \mevcc\
range are fitted using the \piz\ mass constraint, retaining up to 10 candidates.
The presence of at least one well-reconstructed $\gamma_s$ with an energy greater than 50 \mev\ is required. This higher threshold for $\gamma_s$ is motivated by signal MC simulations but also 
reduces background from random soft $\gamma$'s originating from final-state radiation effects and beam background.

In the following, particle momenta are always evaluated in the laboratory frame.
Signal MC simulations of reaction~(\ref{eq:sig}) indicate that the four charged tracks have 
well-defined kinematics, with the two lepton candidates from \OneS decay each having a momentum greater than 2.9 \gevc\ 
and the two pion candidates from the $\omega$ decay each having 
a momentum less than \mbox{0.7 \gevc}.
Muons, electrons, and pions can be identified by applying high-efficiency particle identification
 criteria~\cite{BaBar:2014omp}. However, for this analysis,
 to minimize systematic uncertainties, it is assumed that the two tracks with the largest momenta are leptons and that the other two are pions.

Particle identification is only used to tag the events as belonging to the $\OneS \to \mumu$ or $\OneS \to\ee$ decay mode. To test the performance of the tagging, signal MC events with \OneS\ decaying to \mumu\ or \ee\ are selected using a very loose muon and a very loose electron identification. It is found that the requirement that at least one of the two tracks be positively identified as an electron
correctly separates the two hypotheses: 99.99\% of \mumu\ and 99.94\% of \ee\ events are correctly classified.

To limit the level of combinations in the signal reconstruction, the distributions of the number of reconstructed $\gamma$ and \piz\ candidates in signal MC and data from inclusive \ThreeS\ decays are compared. It is found that an acceptable initial rejection
 of the combinatorial background is achieved by allowing no more than six $\gamma_s$ candidates and no more than five \piz\ candidates.

The MC description of the soft photon background is studied using
the $\ThreeS \to \pip \pim \OneS$ control sample, with $\OneS \to \mumu$. Data and MC simulations from this \ThreeS decay mode are reconstructed with high purity without any requirement on the presence of $\gamma$'s or \piz's.  Good agreement is found between the data and MC simulation  for the distributions of the number of soft reconstructed $\gamma$'s and \piz's as well as for the $\gamma$ energy distribution. Details are given in Appendix~\ref{sec:appendix}.

\subsection{Background suppression} 
\label{sec:back}

Backgrounds from $udsc$ quark pairs 
are expected to be very small because of the $\OneS \to l^+ l^-$ lepton requirement.
There are two significant backgrounds from \ThreeS\ cascades:
\begin{equation}
\begin{array}{lll}
  \ThreeS \to \gamma_1 \chib&&\\  &\almx \to \gamma_2 \TwoS&\\
  & & \almx \to \pip \pim \OneS
  \end{array}
  \label{eq:back1}
\end{equation}
and 
\begin{equation}
\begin{array}{ll}
\ThreeS \to \piz \piz\TwoS&\\ 
&\almx \to \pip \pim \OneS, 
  \label{eq:back2}
  \end{array}
\end{equation}
where, in both processes,  $\OneS \to l^+l^-$.
In cascade~(\ref{eq:back1}) the final state of interest can be produced if a \piz\ is reconstructed from one of the two photons and a
spurious shower in the calorimeter, and in cascade~(\ref{eq:back2}) by the loss of one $\gamma$ in the decays of the two \piz's.

\begin{figure}[!htb]
\begin{center}
  \includegraphics[width=0.49\textwidth]{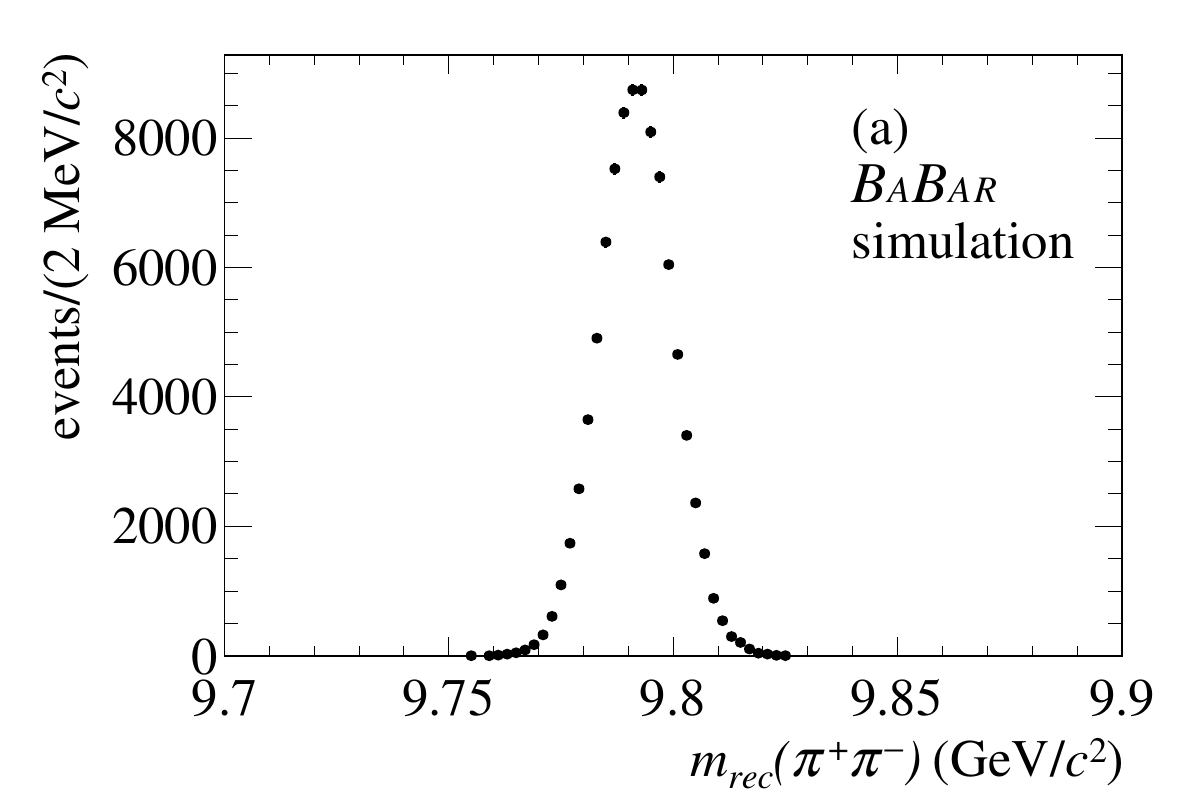}
  \includegraphics[width=0.49\textwidth]{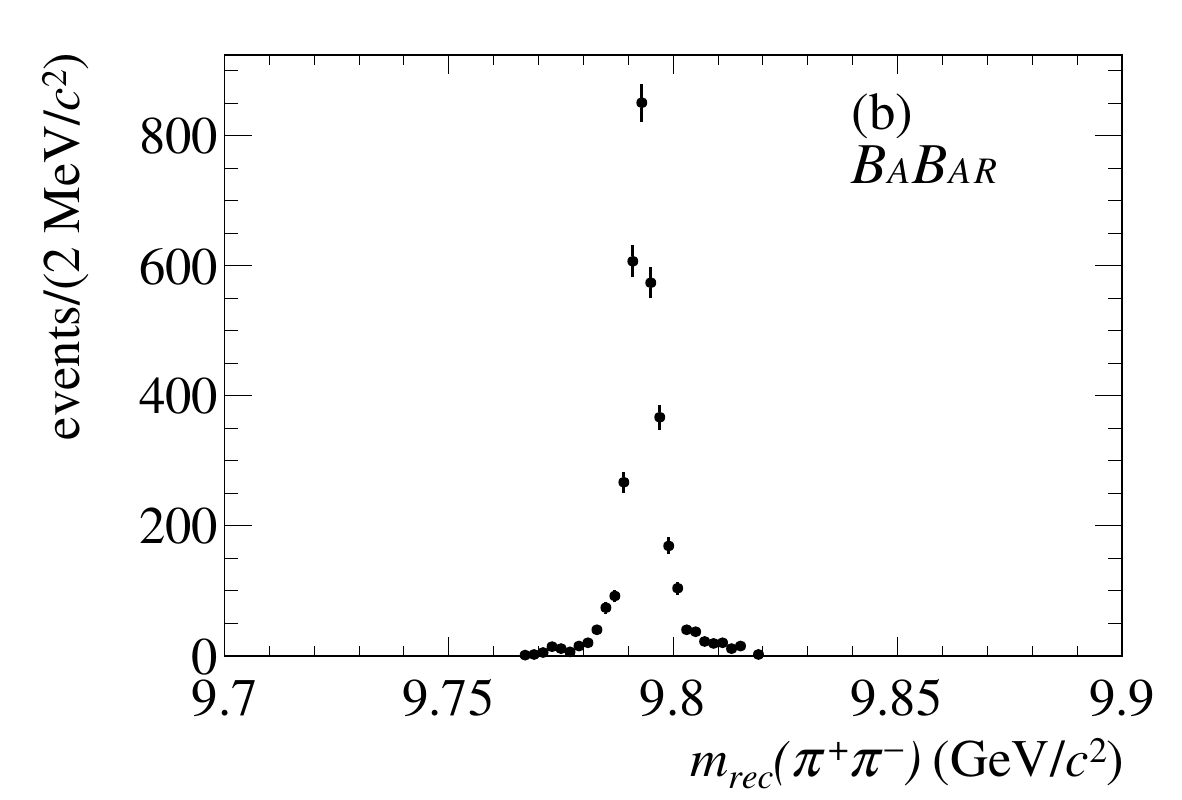}
  \caption{Recoil mass to the $\pip \pim$ system $m_{rec}(\pip \pim)$ for: (a) MC simulation of the decay channel~(\ref{eq:back1}), (b) reconstructed events from data selection of the decay
channel~(\ref{eq:back2}).}
\label{fig:fig1}
\end{center}
\end{figure}

Background channel~(\ref{eq:back1}) is studied using MC simulation with $\chib \equiv \chibtotn$.
To isolate this decay, 
first the \OneS\ signal is selected in the (9.2-9.7) \gevcc\ 
region of the \mumu\ mass spectrum, then the \TwoS\ signal is selected in the (10-10.45) \gevcc\ region of the $\OneS \pip \pim$ mass spectrum.
For each pair of distinct $\gamma_1$ and $\gamma_2$ candidates, the $\Delta m^2$ variable is computed 
\begin{equation}
  \Delta m^2 = m_{rec}^2(\gamma_1) - m^2(\TwoS \gamma_2),
\end{equation}
with
\begin{equation}
  m_{rec}(\gamma_1) = \sqrt{|p(\ThreeS) - p(\gamma_1)|^2}
\end{equation}
and
\begin{equation}
  p(\ThreeS) = p(\ep) + p(\en),
  \label{eq:threes}
\end{equation}
where the symbol $p$ refers to the four-momentum of the particles considered.
Events in background channel~(\ref{eq:back1}) are identified by requiring  $|\Delta m^2|<0.5$ GeV$^2/c^4$. 
For these events, Fig.~\ref{fig:fig1}(a) shows the recoil mass to the $\pip \pim$ system,
\begin{equation}
  m_{rec}(\pip \pim) = \sqrt{|p(\ep) + p(\en) - p(\pip) - p(\pim)|^2}.
  \label{eq:mrec_pipi}
\end{equation}
The background is seen to cluster around a mass of 9.79 \gevcc. 

Background channel~(\ref{eq:back2}) is studied using data where the \TwoS signal is reconstructed as for decay channel~(\ref{eq:back1}). A double loop is then performed over the list of \piz\ candidates, taking care that the two \piz's do not share any $\gamma$.
The variable
\begin{equation}
  \Delta m^2 = m_{rec}^2(\piz \piz) - m^2(\TwoS)
\end{equation}
is then evaluated. The recoil mass $m_{rec}(\pip \pim)$ is shown in Fig.~\ref{fig:fig1}(b) for events inside the squared mass window $|\Delta m^2|<1$ 
GeV$^2/c^4$. 
Similarly to background channel~(\ref{eq:back1}), the events from background channel~(\ref{eq:back2}) are seen to accumulate at values around \mbox{9.79 \gevcc}. 

\begin{figure}[!htb]
\begin{center}
  \includegraphics[width=0.49\textwidth]{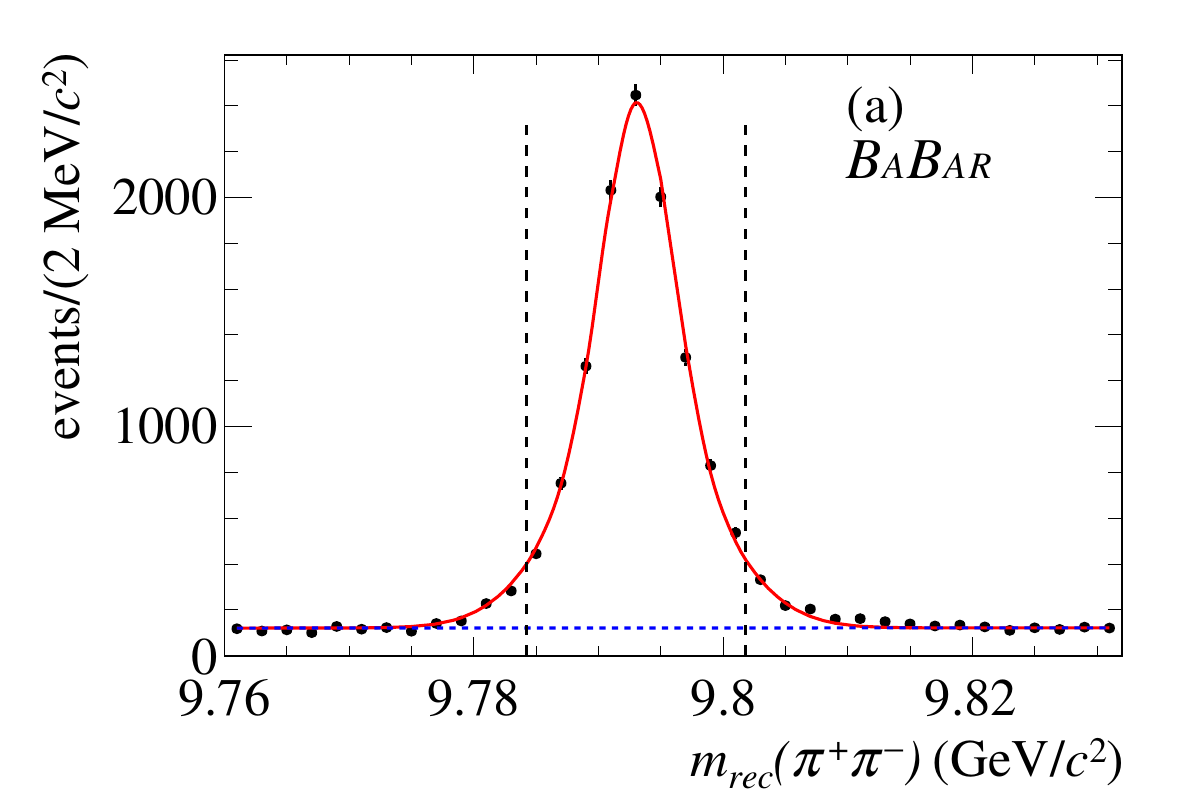}
  \includegraphics[width=0.49\textwidth]{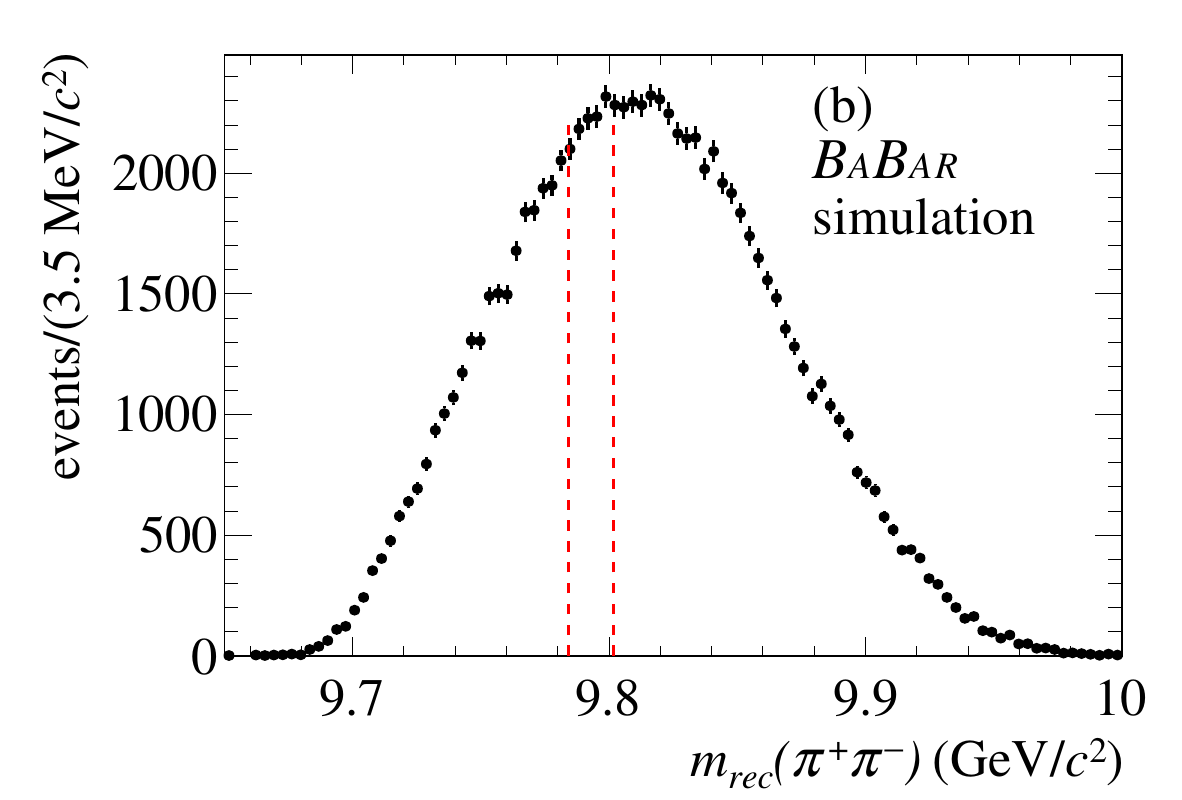}
  \caption{(a) Distribution of the recoil mass against the $\pip \pim$ system, for events selected as candidates for the signal decay channel~(\ref{eq:sig}). The lines are the results from the fit described in the text.
    (b) Recoil mass to the $\pip \pim$ system for signal MC decay channel~(\ref{eq:sig}) events. The vertical lines indicate the region removed by the veto on background events from channels~(\ref{eq:back1}) and~(\ref{eq:back2}).}
  \label{fig:fig2}
\end{center}
\end{figure}

Figure~\ref{fig:fig2}(a) shows the $m_{rec}(\pip \pim)$ distribution for all signal candidate events in the decay channel~(\ref{eq:sig}) for the mass region around 9.79 \gevcc.
The expected narrow structure from the background decay channels~(\ref{eq:back1}) and (\ref{eq:back2}) is observed.

The $m_{rec}(\pip \pim)$ distribution shown in Fig.~\ref{fig:fig2}(a) is fitted using two Gaussian functions sharing the same mean on top of a linear background.
Fitted parameters are \mbox{$m=(9.7931\pm0.0001)$ \gevcc} and \mbox{$\sigma_e=(4.4 \pm 0.4)$ \mevcc}.
Here and below, in fits performed using two Gaussian functions, the effective $\sigma_e$ is defined as
\begin{equation}
  \sigma_e = f\sigma_1 + (1 - f) \sigma_2
  \label{eq:sigma}
\end{equation}
where $\sigma_1$ and $\sigma_2$ are the two fitted Gaussian widths and $f$ is the fitted fraction of the first Gaussian contribution.

A veto is applied to events having $m_{rec}(\pip \pim)$ in the $m \pm 2\sigma_e$ mass region.
Figure~\ref{fig:fig2}(b) shows the $m_{rec}(\pip \pim)$ distribution from MC for signal channel~(\ref{eq:sig}) events. The region of the veto, which eliminates around 12\% of the signal events, is indicated.
The background channels (\ref{eq:back1}) and (\ref{eq:back2}) have branching fractions, for the \chibone case,  of $(3.23 \pm 0.34)\%$ and $(0.33 \pm 0.03)\%$, respectively.
It is found that the combination of the $m_{rec}(\pip \pim)$ veto and the criteria used for selecting the signal (\ref{eq:sig}), described in the following, suppresses the two background channels by a factor $\approx 2\times 10^{-3}$ and therefore their contributions become negligible. After vetoing the background channels, events are further processed through the event selection procedure. 

\subsection{Event selection} 
\label{sec:event}

The data selection begins with a loop over $\gamma_s$ candidates,
  followed by a loop over \piz\ candidates excluding \piz's reconstructed with the $\gamma_s$.
Momentum balance is required, using 
the distributions of
$\Delta {\it p}_i$, $i\equiv x,y,z$,  
the missing three-momentum 
components of the initial and final-state particles
\begin{equation}
\Delta {\it p}_i = {\it p}_i(e^+) + {\it p}_i(e^-) - \sum_{j=1}^{6}{\it p}_i(j).
\label{eq:chi1}
\end{equation}
In Eq.(\ref{eq:chi1}), ${\it p}_i$
indicates the three components of the
laboratory momenta of the six particles in the final state and of the two
incident beams, with the $z$ axis along the beam direction.
The $\Delta {\it p}_i$ distributions are evaluated for data and signal MC
simulations and are fitted using two or
three Gaussian functions centered at zero.
When multiple Gaussian functions are used, the quoted widths
$\sigma$ are the effective $\sigma_e$ evaluated using Eq.~(\ref{eq:sigma}). The fitted values are $\sigma_{xy}=75$ \mevc\ 
and $\sigma_z=85$ \mevc. Combinations within $\pm 3\sigma$  are selected.

\begin{figure}[!htb]
\begin{center}
  \includegraphics[width=0.49\textwidth]{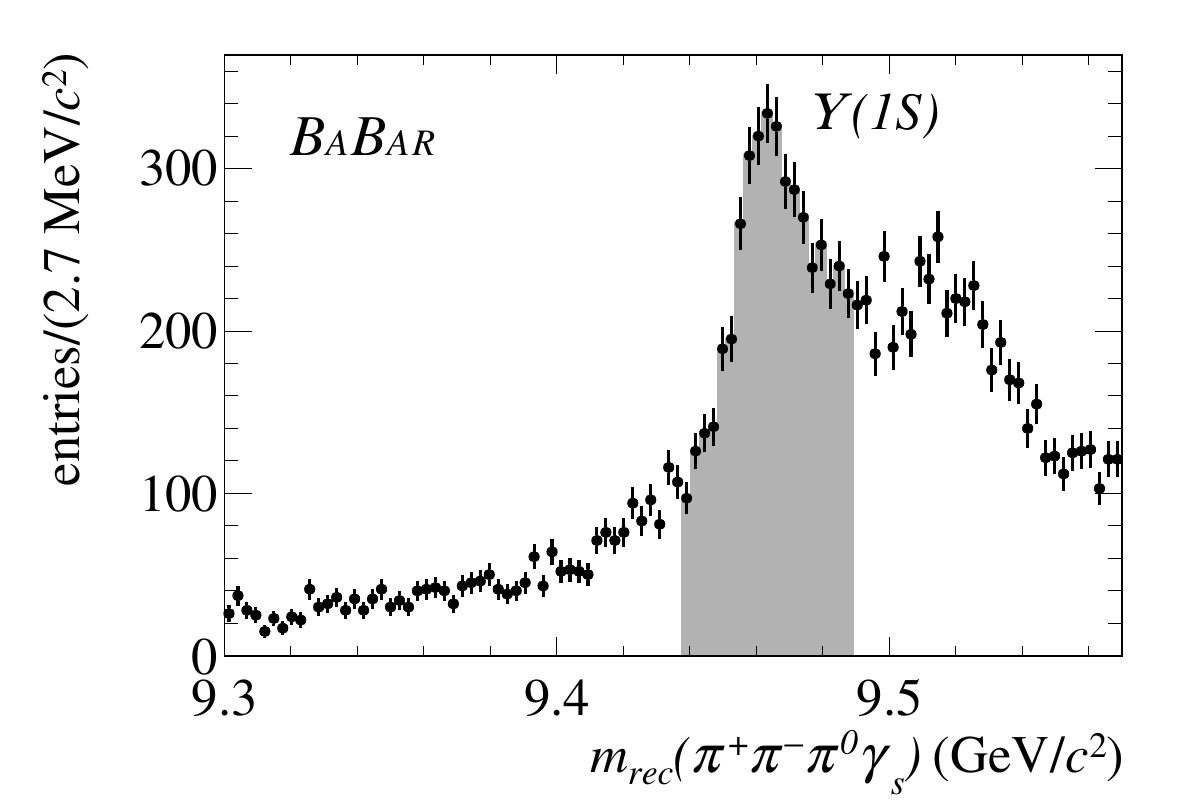}
  \caption{Combinatorial recoil mass $m_{rec}(\pip \pim \piz \gamma_s)$. The shaded area 
shows the region of selected \OneS candidates. The enhancement above the \OneS mass region is produced by the combinatorial.}
  \label{fig:fig3}
\end{center}
\end{figure}

\begin{figure}[!htb]
\begin{center}
  \includegraphics[width=0.49\textwidth]{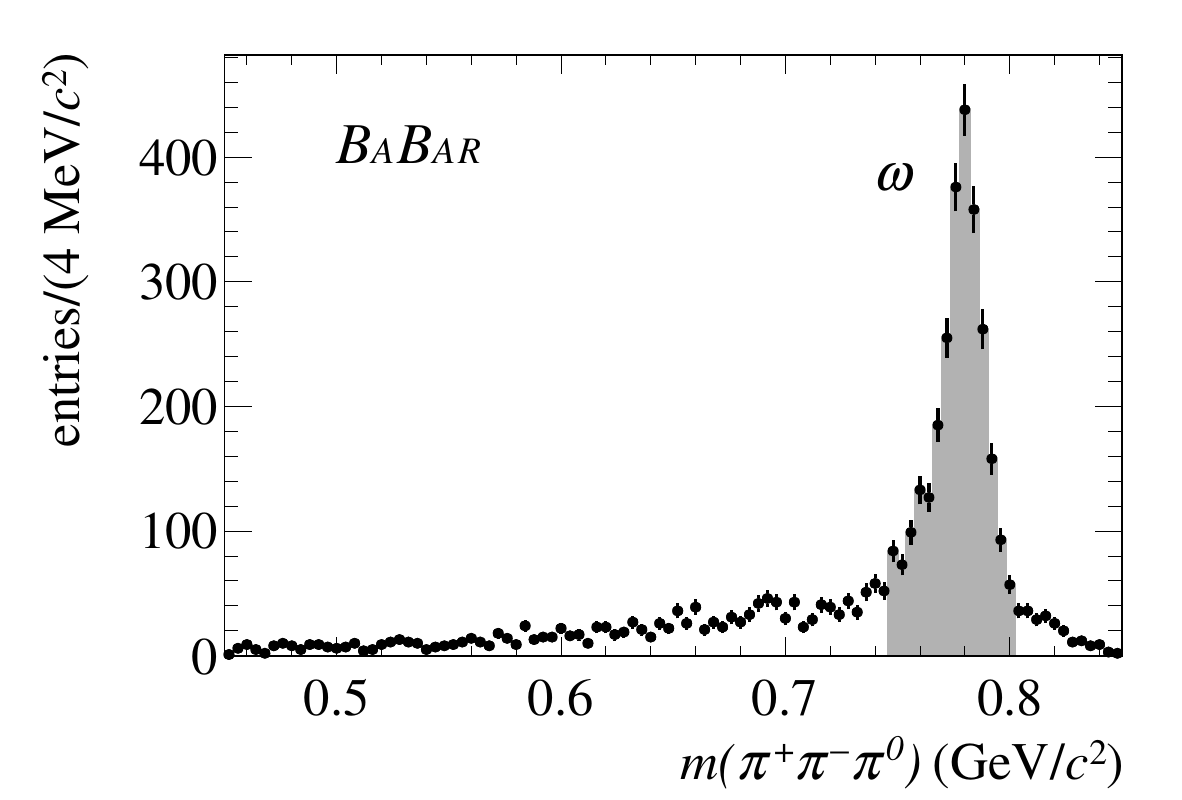}
  \caption{The $\pip \pim \piz$ invariant mass distribution. The shaded area indicates the range of the $\omega$ selection.} 
  \label{fig:fig4}
\end{center}
\end{figure}

Figure~\ref{fig:fig3} shows the recoil mass $m_{rec}(\pip \pim \piz \gamma_s)$ 
\begin{equation}
  m_{rec}(\pip \pim \piz \gamma_s) = \sqrt{|p(\ThreeS) - p(\pip \pim \piz \gamma_s)|^2},
\end{equation}
where $p$ indicates
the 4-momenta of the initial- and final-state particles. The mass spectrum is dominated by the \OneS\ signal on top of a significant combinatorial background.
Due to the large combinatorial and peaking backgrounds, the \OneS\ parameters are obtained from signal MC for which the combinatorial background is suppressed using MC information (MC truth). The fit yields 
\mbox{$m(\OneS)=9.4633$ \gevcc}\ and 
$\sigma(\OneS)=8.4$ \mevcc. 
The \OneS\ signal is selected within $\pm 3\sigma$.
The resulting $\pip \pim \piz$ mass spectrum is shown in Fig.~\ref{fig:fig4}. It is dominated by the $\omega$ signal,
which is selected in the range \mbox{$0.75<m(\pip \pim \piz)<0.8$ \gevcc.} 

The reconstruction of the signal is completed by adding the 
two candidate lepton 
three-vectors and computing the \OneS\ energy using its known mass~\cite{PDG}. Similarly, the $\omega$ four-vector is obtained by adding the \pip, \pim, and \piz\ three-momenta and computing the $\omega$ energy using its known mass.

The procedure is tested using MC simulations from the decay channel Eq.~(\ref{eq:back1}).
  The invariant mass resolution in the region of the \TwoS is measured as $\sigma_e(\TwoS \to \OneS \pip \pim)=67.9 \pm 2.8$ \mevcc. A significant improvement of the \TwoS mass resolution, \mbox{$\sigma_e=2.7 \pm 0.1$ \mevcc},  is obtained by adding the two leptons momenta from the \OneS decay and computing its energy using its known mass.
  
This method has been found to significantly improve invariant mass resolutions in charmonium decays involving $\eta$, \etapr~\cite{BaBar:2021fkz} or \KS decays~\cite{LHCb:2023evz}, where an approximately 6\% worse resolution was estimated with respect to that obtained by a mass constrained fit.
In the present analysis MC simulations show that the effect of the procedure on the $\omega$ reconstruction is to symmetrize the distribution of the $\itdelta^2$ variable defined below.

\begin{figure}[!htb]
\begin{center}
  \includegraphics[width=0.49\textwidth]{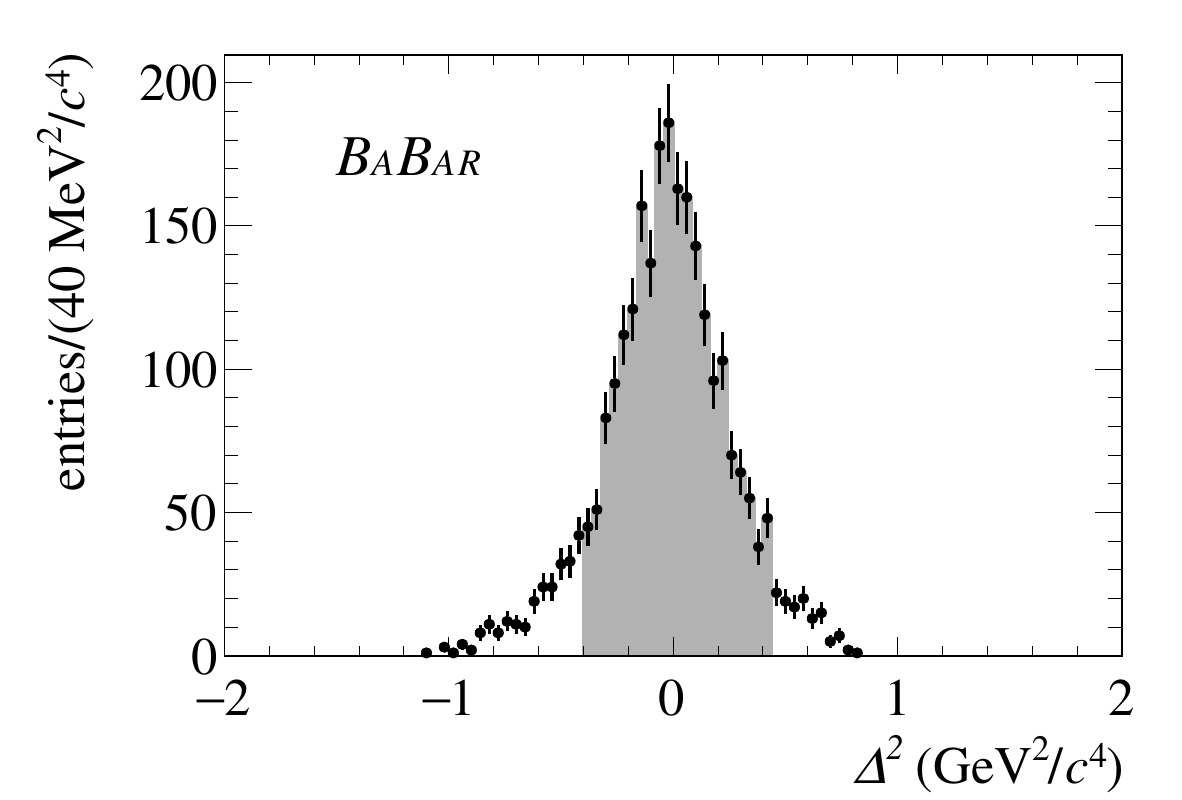}
  \caption{Distribution of $\itdelta^2$ for the candidate events. The shaded area indicates the range of the final selection.} 
  \label{fig:fig5}
\end{center}
\end{figure}

The variable $\itdelta^2$ is then evaluated as

\begin{equation}
  \itdelta^2 = m_{rec}^2(\gamma_s) - m^2(\OneS\ \omega)
  \label{eq:delta}
\end{equation}
and is shown in Fig.~\ref{fig:fig5}. 

The $\itdelta^2$ variable is used to (a) remove background combinations having $|\itdelta^2|>\itdelta^2_{lim}$ and (b) select the candidate having the lowest $|\itdelta^2|$ value in the case of two combinations. Using signal MC events, $\itdelta^2_{lim}$=0.4 GeV$^2/c^4$ is selected as the optimal value to reduce the combinatorial background without affecting the size of the \chibtot signals. 
The selected interval is shown as the shaded region in Fig.~\ref{fig:fig5}. For this selection, 78.8\%, 17.6\%, and 3.6\% of candidates have one, two,  or more than two combinations, respectively. Events having more than two combinations are discarded.

As stated in the beginning of Sec.~\ref{sec:event}, the \ThreeS decay chain under study contains two high and two low momentum charged tracks in addition to three soft photons. It is found that, in this particular kinematical configuration, invariant mass resolutions are rather poor.
  Using signal MC simulation, resolution functions are evaluated as the differences
  between the MC-generated and reconstructed invariant masses. The resulting distributions are fitted by the sum of two Gaussian functions. The resulting $\sigma_e$~(Eq.(\ref{eq:sigma})) is found to be \mbox{$5.8 \pm 0.2$ \mevcc} for the $\omega$ (to be compared with the known width of $8.68\pm0.13$ \mevcc~\cite{PDG}) and $68.0 \pm 2.9$ \mevcc\ for the $\OneS \to \ellp \ellm$ invariant mass.
  
For these reasons, the established procedure to obtain the \chib lineshapes is to evaluate the distribution of the center-of-mass $\gamma_s$ momentum in the $\Upsilon$ decays~\cite{BaBar:2014och,PDG}. Due to the well measured $\epem$ beam energies, the resulting resolution depends only on the
  measured $\gamma_s$ energy. In the present analysis the recoil mass $m_{rec}(\gamma_s)$ is used, defined below, which is obtained from the same variables used by the center-of-mass $\gamma_s$ momentum and therefore has the same resolution.

\begin{figure}[!htb]
\begin{center}
  \includegraphics[width=0.49\textwidth]{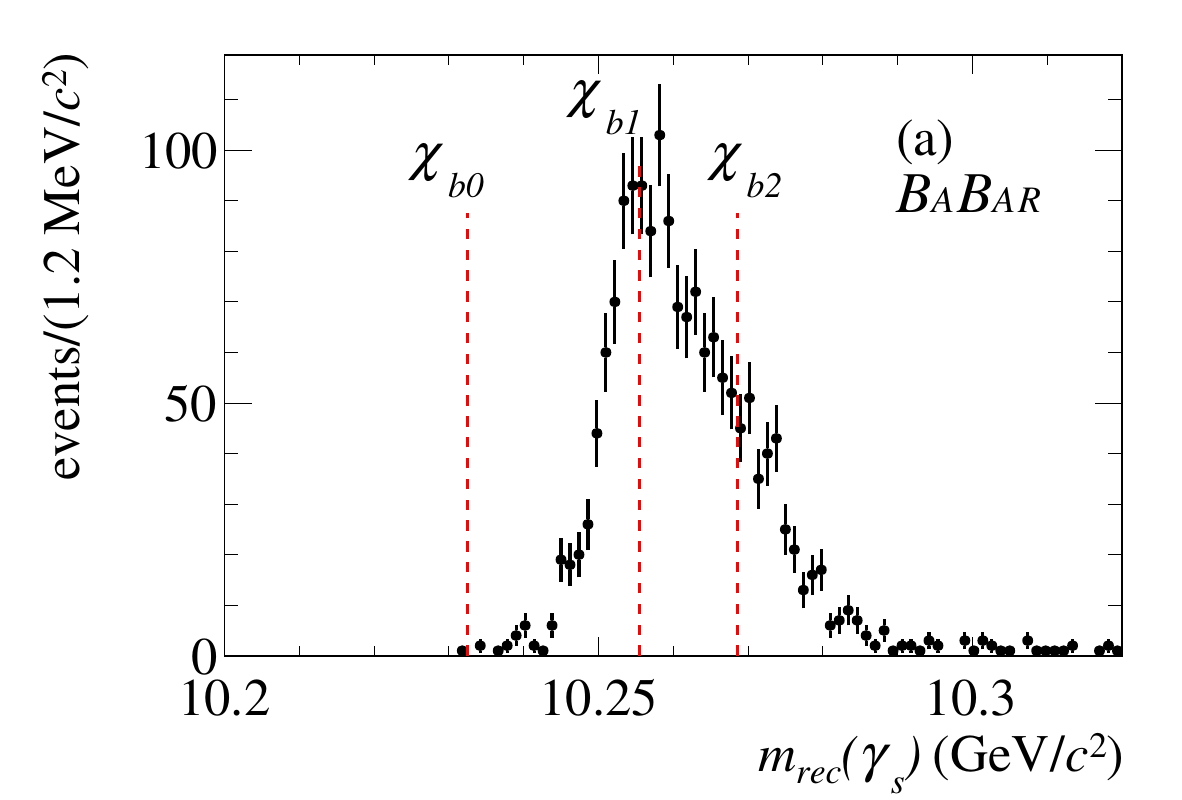}
  \includegraphics[width=0.49\textwidth]{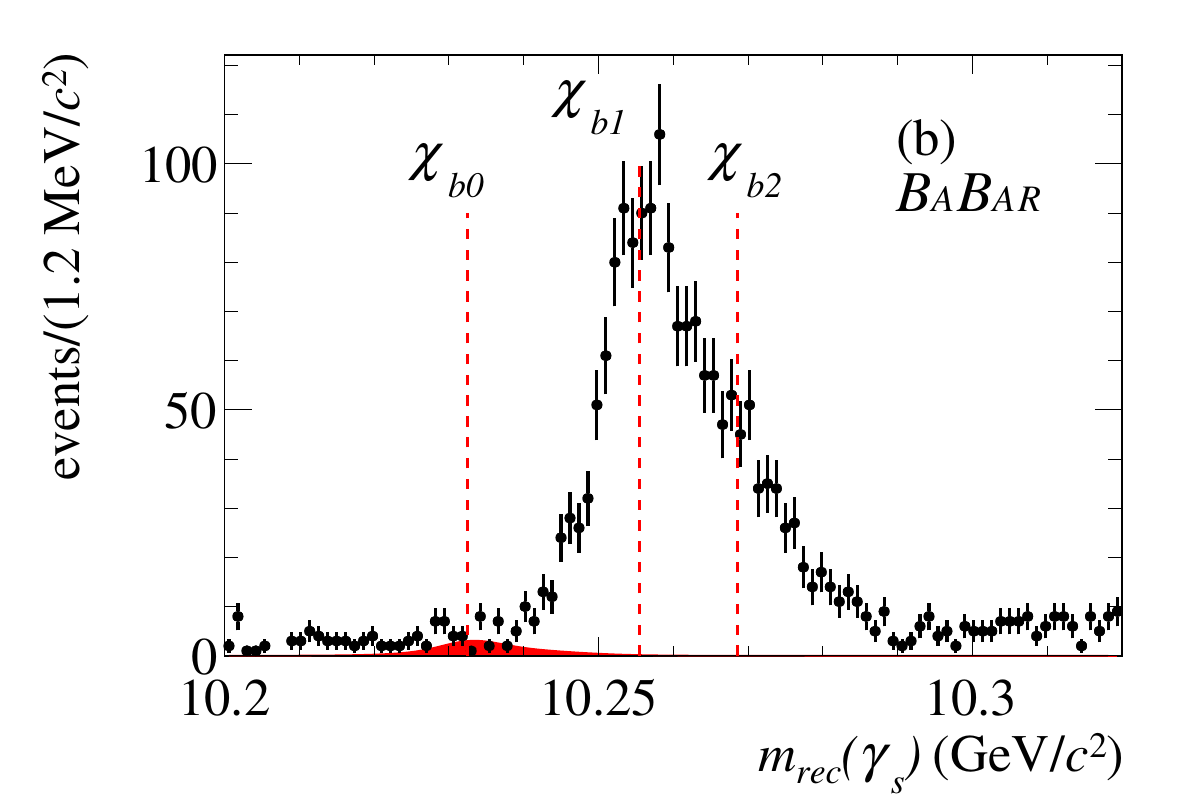}
  \caption{Distribution of $m_{rec}(\gamma_s)$ for candidate events (a) after all selections and (b) removing the mass selection on the $\omega$.
    The vertical lines indicate the mass of the \chibtotn states. The small shaded (red) distribution indicates the estimated 90\% upper limit contribution (41 events) from the \chibzero resonance.} 
  \label{fig:fig6}
\end{center}
\end{figure}

Figure~\ref{fig:fig6}(a) shows the $m_{rec}(\gamma_s)$ distribution, defined by
\begin{equation}
  m_{rec}(\gamma_s) = \sqrt{|p(\ThreeS) - p(\gamma_s)|^2},
  \label{eq:mrec_g}
\end{equation}
for the 1651 selected candidates (81\% \mumu and 19\% \ee).

In the \babar\ dataset, trigger-level prescaling of Bhabha events is employed.
  These events are characterized by two electrons of large invariant mass and no
additional charged track with transverse 
momentum $> 250$ \mevc. The prescaling causes the efficiency to be smaller for events with electrons than for events with muons.  
Therefore, since the efficiency dependence on angular variables for both samples is similar,  the combined dataset is used to
measure the \chibtot angular distributions, but only the \mumu data are
used to evaluate the branching fractions. This choice is motivated by the small size of the \ee dataset and the larger systematic uncertainties associated with the simulation of the trigger and the electron reconstruction.

Overlapping \chibone and \chibtwo signals are observed, while the possible presence of a \chibzero signal is strongly suppressed by the $\omega \OneS$ invariant mass threshold.
To test for the possible presence of a \chibzero arising from the $\omega$ low-mass tail, the selection on the $\pip \pim \piz$ invariant mass is removed, as well as
the assignment of the known $\omega$ mass to the sum of the three-pion  
system. The resulting $m_{rec}(\gamma_s)$ distribution is shown
in Fig.~\ref{fig:fig6}(b). Again, there is no evidence for a \chibzero signal.

\subsection{Efficiency}
\label{sec:effy}

To compute the efficiency, signal MC events from cascade~(\ref{eq:sig}) are generated using a
detailed detector simulation~\cite{geant}. The kinematics of the simulation are governed by phase space~\cite{Lange:2001uf} except for the expected spin-one $\omega$ angular distribution.
These simulated events are reconstructed and analyzed in the same manner as data. The efficiency is then computed as the ratio between 
reconstructed and generated events and projected over several kinematic variables.

The following angles are defined.

\begin{figure}[!htb]
\begin{center}
  \includegraphics[width=0.40\textwidth]{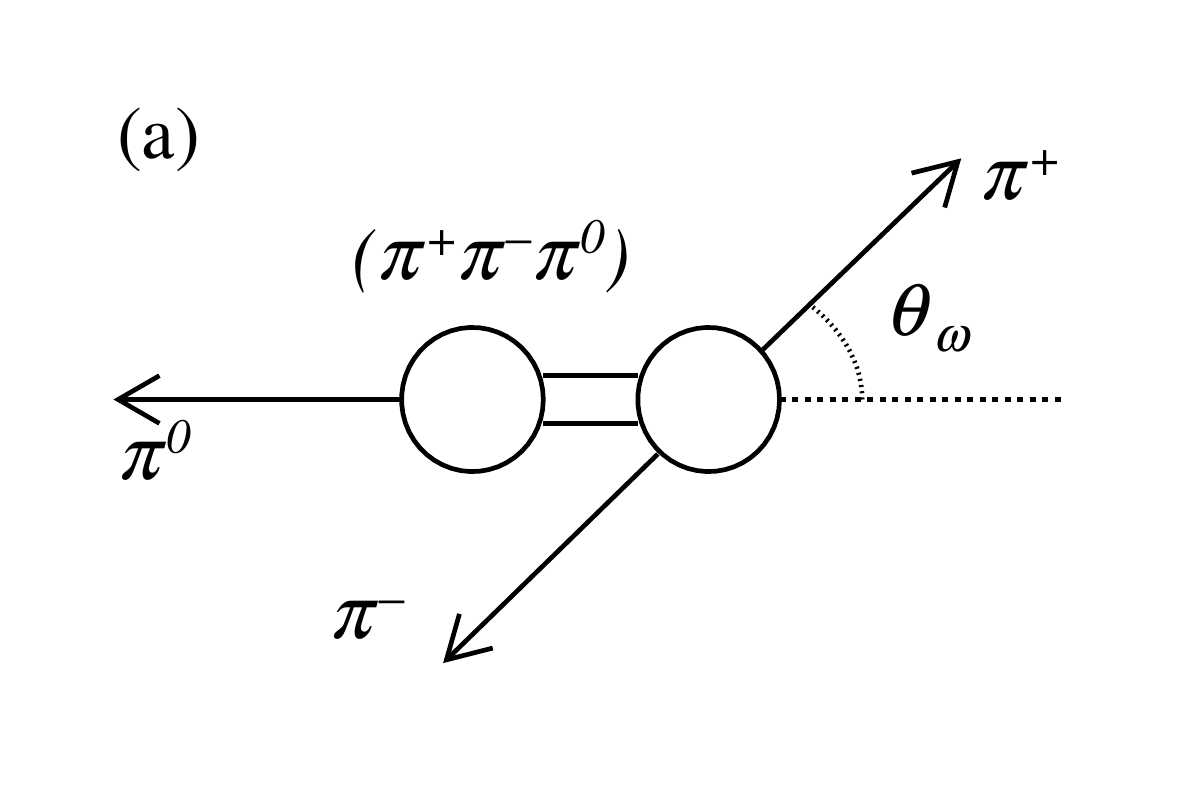}
  \includegraphics[width=0.49\textwidth]{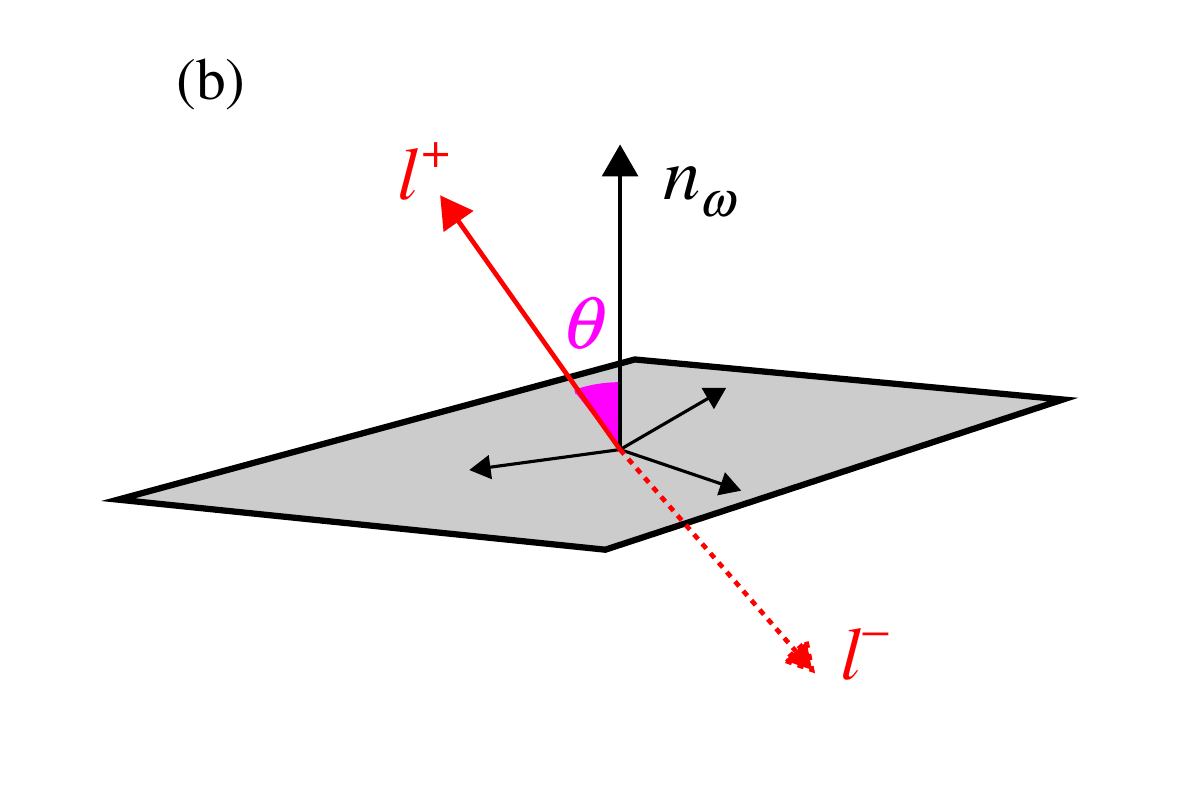}
  \includegraphics[width=0.40\textwidth]{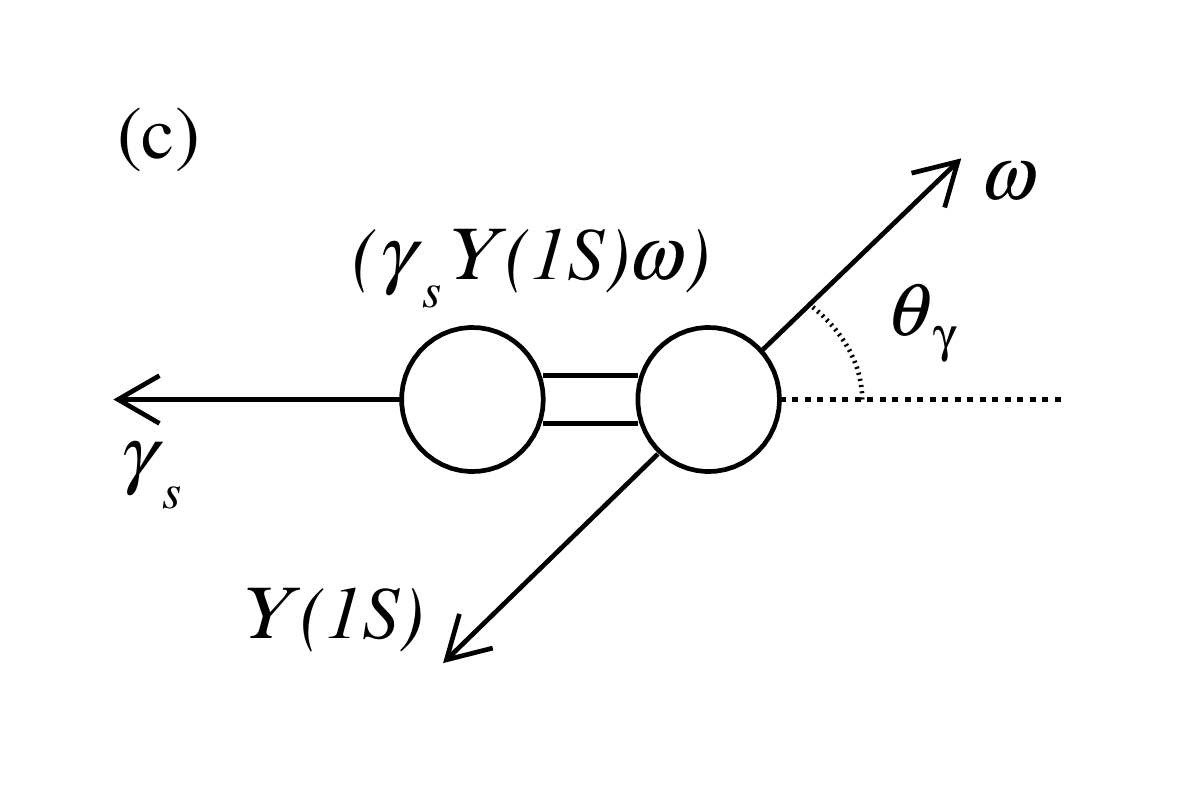}
\caption{Definition of the angles $\theta_{\omega}$, $\theta$, and $\theta_{\gamma}$ describing the decay $\ThreeS \to \gamma_s\omega\OneS$. In (b), for simplicity, the $\omega$ is assumed to be at rest in the \OneS\ rest frame.}
 \label{fig:fig7}
\end{center}
\end{figure}

\begin{itemize}
\item{(a)} In the $\omega$ signal region, $\theta_{\omega}$ is defined as the angle, in the $\pip \pim$ rest frame, between the direction of the \pip\ and the boost from the $\pip \pim$ system (Fig.~\ref{fig:fig7}(a)).

\item{(b)} In the $\OneS$ rest frame, $\theta$ is defined as the angle formed by one of the two leptons from the \OneS\ decay with the normal $n_{\omega}$ to the $\omega$ plane boosted in the \OneS\ rest frame (Fig.~\ref{fig:fig7}(b)).

\item{(c)} In the $\gamma_s \OneS \omega$ signal region, $\theta_{\gamma}$ is defined as the angle, in the $\OneS \omega$ rest frame, between the direction of the $\omega$ and the boost from the $\OneS \omega$ system (Fig.~\ref{fig:fig7}(c)).
\end{itemize}

\begin{figure}[!htb]
\begin{center}
  \includegraphics[width=0.49\textwidth]{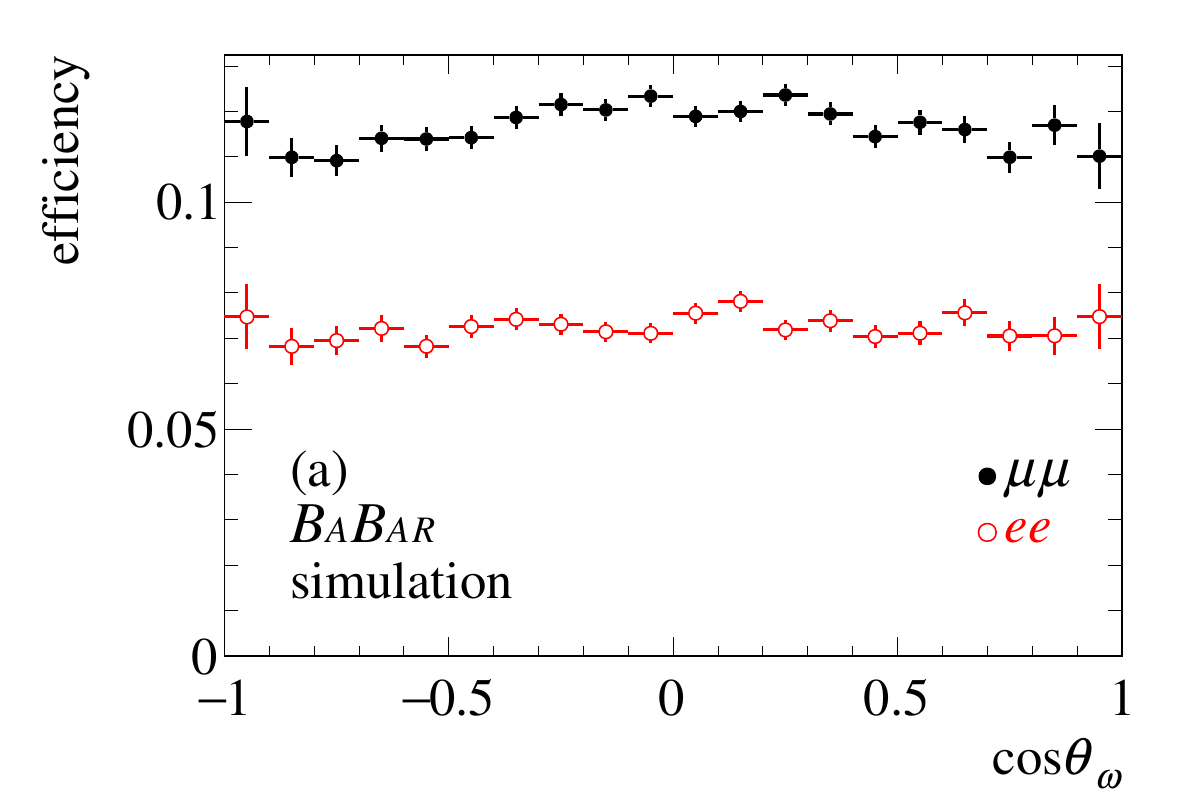}
  \includegraphics[width=0.49\textwidth]{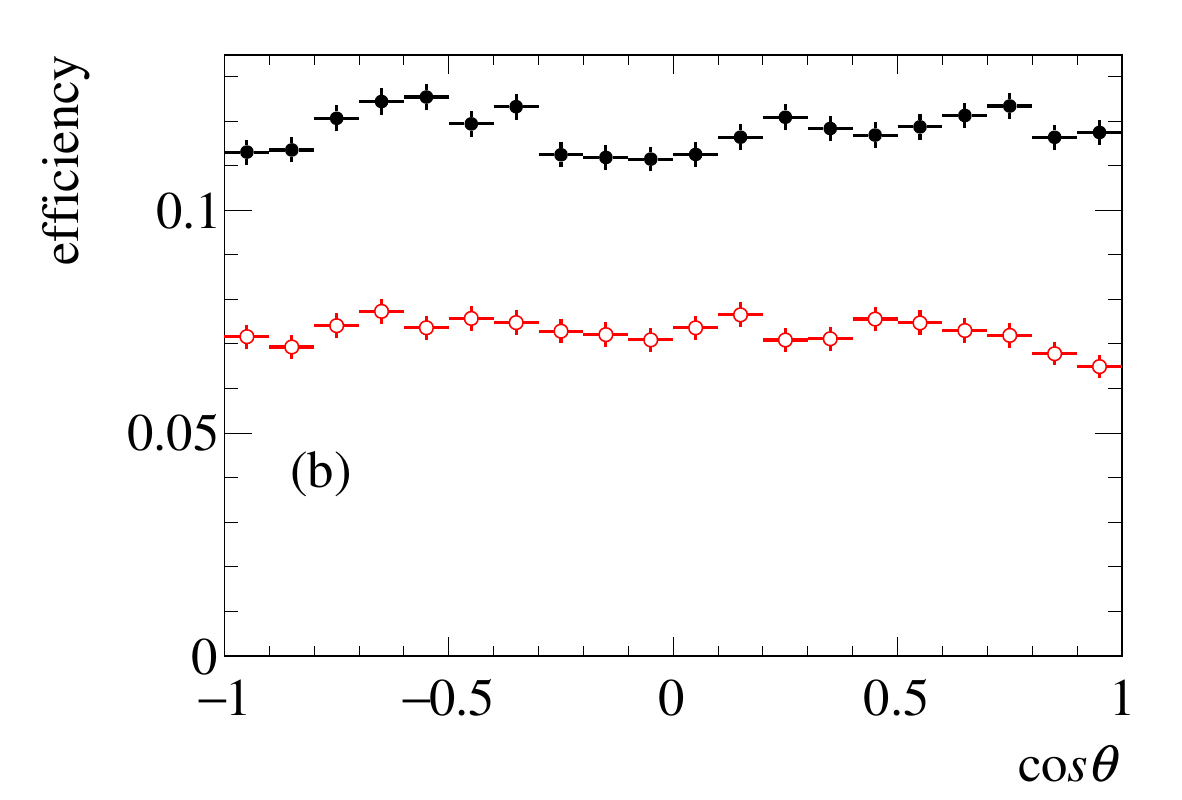}
  \includegraphics[width=0.49\textwidth]{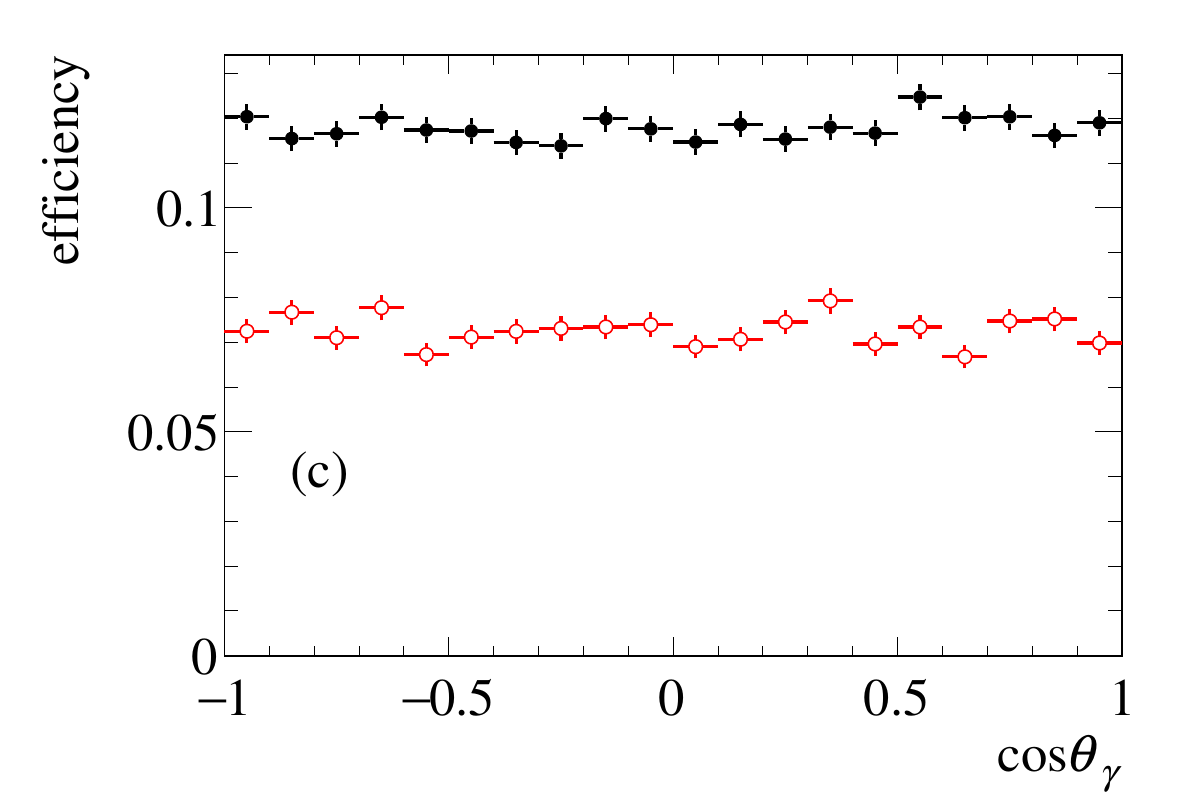}
\caption{Efficiency projections over $\cos\theta_{\omega}$, $\cos\theta$, and $\cos\theta_{\gamma}$, separately for \mumu, and \ee simulations.}
  \label{fig:fig8}
\end{center}
\end{figure}

Figure~\ref{fig:fig8} shows the efficiency projected onto the three
angles, separately for the \mumu and \ee simulations. 
A weak dependence on the three angles can be seen, with the \ee efficiency
significantly smaller than that for \mumu.

\section{Data analysis}

\subsection{Measurement of the \boldmath{\chibtot} fractions}
\label{sec:chan}

To measure the \chibone and \chibtwo fractional contributions in the selected data, described by the $m_{rec}(\gamma_s)$ variable, the method of
Channel Likelihood~\cite{Condon:1974rh} is used.
An unbinned maximum likelihood fit to the data is performed, using an iterative procedure.
The convergence criterion  
is defined by the difference in the fractions between the last two iterations and can be set to an arbitrary value (typically $10^{-5}$ after about 25 iterations).
The total pdf (\calL) is described by
\begin{equation}
  \calL = c_1\cdot f(\chibone) + c_2\cdot f(\chibtwo) + ps
  \label{eq:chan}
  \end{equation}
where $c_1$ and $c_2$ are the fractional contributions for \chibone and \chibtwo. The symbol $ps$ indicates the residual contribution (in this case the background), normalized as
$ps = (1 - c_1 - c_2)\cdot b$ with the factor $b$ fixed to a constant value ($b=1$). The symbols $f(\chibone)$ and $f(\chibtwo)$ indicate the
functional lineshapes describing the \chibone and \chibtwo contributions, integrated over the $m_{rec}(\gamma_s)$ fit range. The method assigns, after the fit, weights $w(\chibone)$, $w(\chibtwo)$, and $w(ps)$ to each event for it to belong to the different components included in the definition of the pdf function.
Examples of the use of the channel likelihood method can be found in Refs.~\cite{JETSET:1998akg,BaBar:2003cdx}.

    The \chibtot functional lineshapes are described by the Breit-Wigner functions ({\it BW})
    \begin{equation}
      BW(m) = \frac{1}{(m_0 - m)^2 +\itgamma^2/4}
      \label{eq:bw}
      \end{equation}
    convolved with the detector resolution functions and the $m_0$ values fixed to known values for \chibtot~\cite{PDG}.
The $\chi_{bJ}(2P)$ total widths are not experimentally measured, but are expected to be small.
  Reference~\cite{Godfrey:2015dia} predicts values of 232.5 keV, 117.0 keV
and 2.6 MeV for the $\chi_{b2,b1,b0}(2P)$ states, respectively.    
    The value of \itgamma is set to an arbitrary value of \itgamma=0.1 \mev\, in the range of the small expected widths of these states. 

The resolution functions $\delta(m_{rec}(\gamma_s))$ are
evaluated separately for the \chibone and \chibtwo as the difference
between the MC-generated and reconstructed $m_{rec}(\gamma_s)$.
To remove   
multiple combinations, MC truth information is used.
The distributions are fitted by the sum of a Gaussian function ($G$, c=57\%, with $\sigma=7.5$ \mevcc)  
and a Crystal-Ball function ($CB$, with $\sigma=3$ \mevcc),  
\begin{equation}
  BW_c(x) = BW(x)\otimes[(1-c)\cdot CB(x) + c \cdot G(x)],
  \label{eq:bw_conv}
\end{equation}
similarly for both \chib states.
The fitting model is validated on signal \mumu MC simulations by performing fits with known \chibtot contributions. Good agreement is found, within the uncertainties, between the input and fitted fractional contribution values.
The results from the likelihood fits to the simulated samples are listed in Table~\ref{tab:tab1}, which summarizes the generated and reconstructed yields for the \chibone and \chibtwo together with the resulting efficiencies, separated 
for \chibone and \chibtwo and for the \mumu and \ee simulations.

\begin{table} [!htb]
  \caption{Results from the fits to signal MC events. The numbers of generated and reconstructed events
  and the corresponding efficiencies for the separated \chibone and \chibtwo states, and the combined
  efficiencies, are reported for the \mumu and \ee channels.}
    \label{tab:tab1}  
    \begin{center}
\begin{tabular}{lcc}
  \hline\\ [-2.3ex]
      & \chibone & \chibtwo \cr
  \hline\\ [-2.3ex]
  & \alx \mumu & \cr
  \hline\\ [-2.3ex]
  Generated &191959 & 128443   \cr
  Reconstructed & $22722 \pm 171$ & $14399 \pm 146$ \cr
  Efficiency (\%)& \al $11.84\pm  0.09$ & \al$11.21 \pm  0.12$ \cr 
    Average (\%)& \alx$11.60 \pm  0.07$  &   \cr
    \hline\\ [-2.3ex]
      & \alx \ee & \cr
    \hline\\ [-2.3ex]
  Generated & 125342 & 84175 \cr
  Reconstructed &$9184 \pm 109$  & $5719 \pm 92$ \cr
  Efficiency (\%) &  \al $7.33 \pm  0.09$ &  \al $6.79 \pm  0.11$ \cr
  Average (\%)&  \alx $7.12 \pm   0.07 $ & \cr
\hline
\end{tabular}
\end{center}
\end{table}

The summed 
\mumu and \ee data are first processed through the channel likelihood fit setting \itgamma=0.1 \mev\ in Eq.~(\ref{eq:bw}) for both \chibone and \chibtwo.
The fit quality is assessed by evaluating 
the $\chisq$/ndf  
value for the \mrec distribution
by comparing the data with the fitted function given by Eq.~(\ref{eq:chan}).

\begin{figure}[!htb]
\begin{center}
  \includegraphics[width=0.49\textwidth]{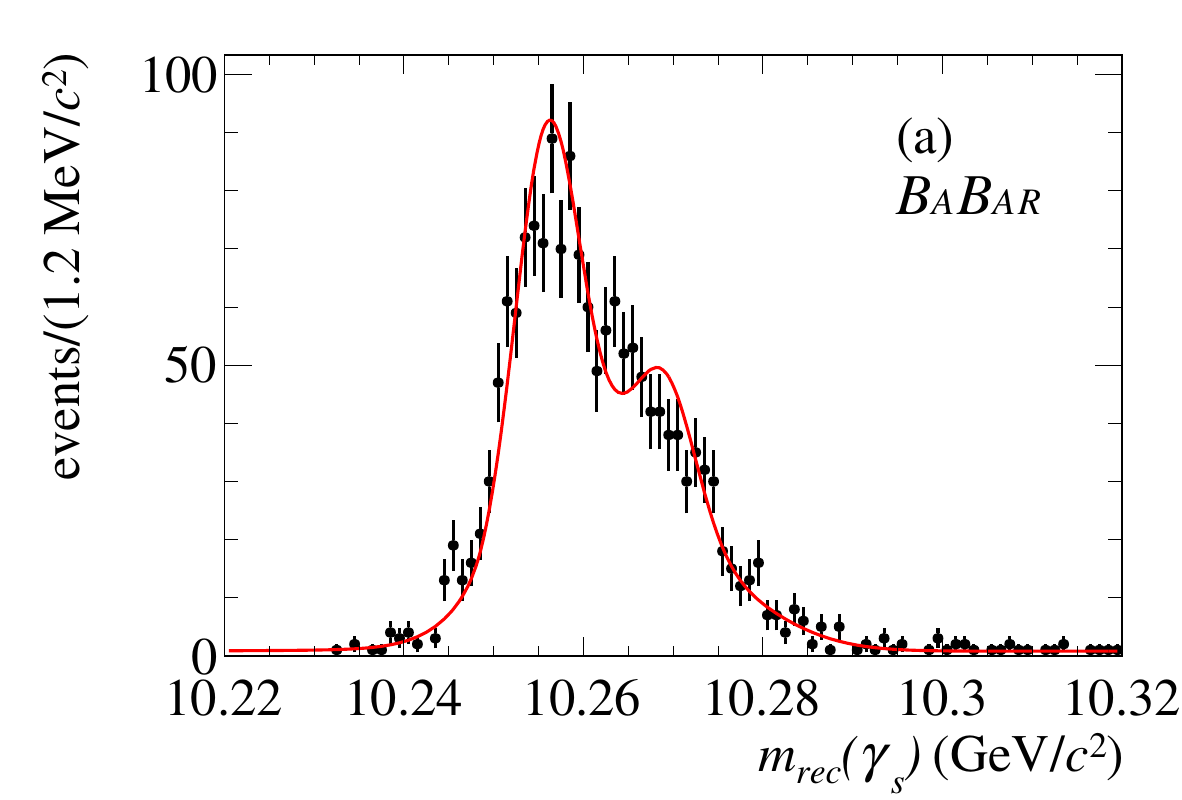}
  \includegraphics[width=0.49\textwidth]{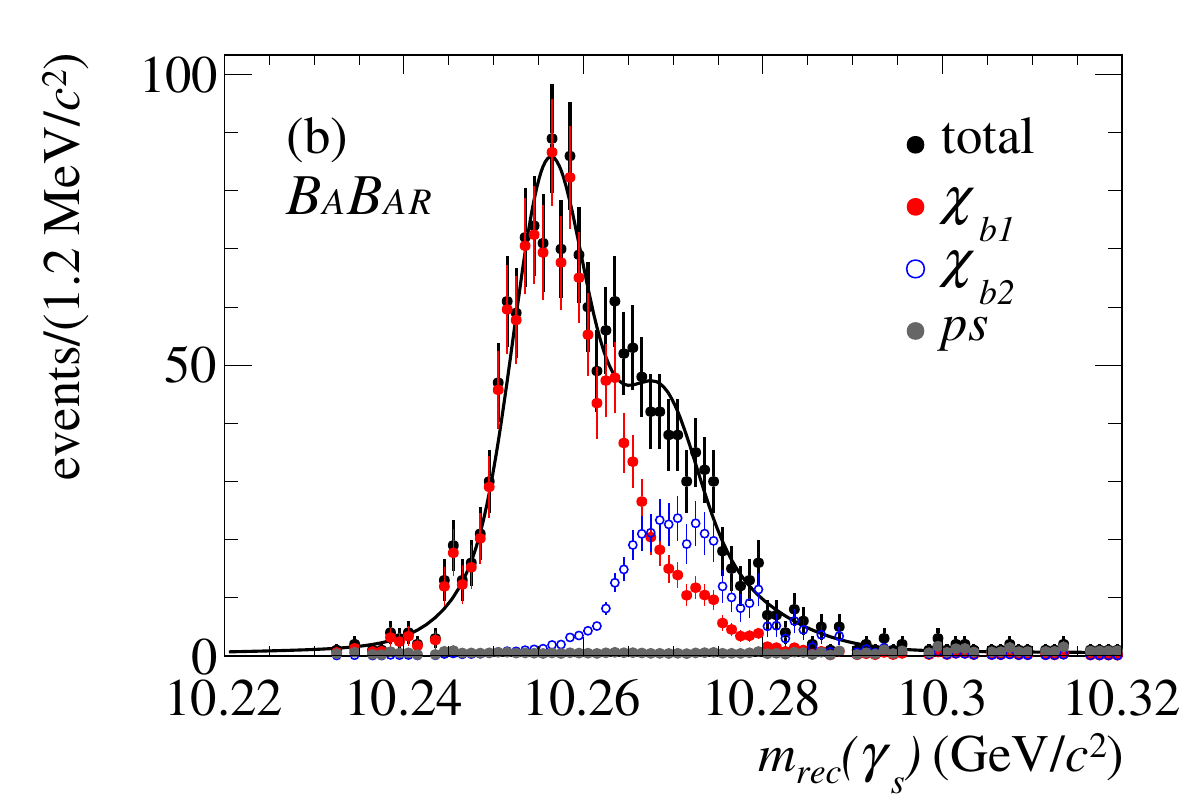}
\caption{Fit projections on the \mrec distribution of the total data sample. The total fitting functions for (a) \mbox{\itgamma=0.1 MeV} and (b) \itgamma=1.4 \mev\ with the \chibtot components are superimposed.}
  \label{fig:fig9}
\end{center}
\end{figure}

\begin{figure}[!htb]
\begin{center}
  \includegraphics[width=0.49\textwidth]{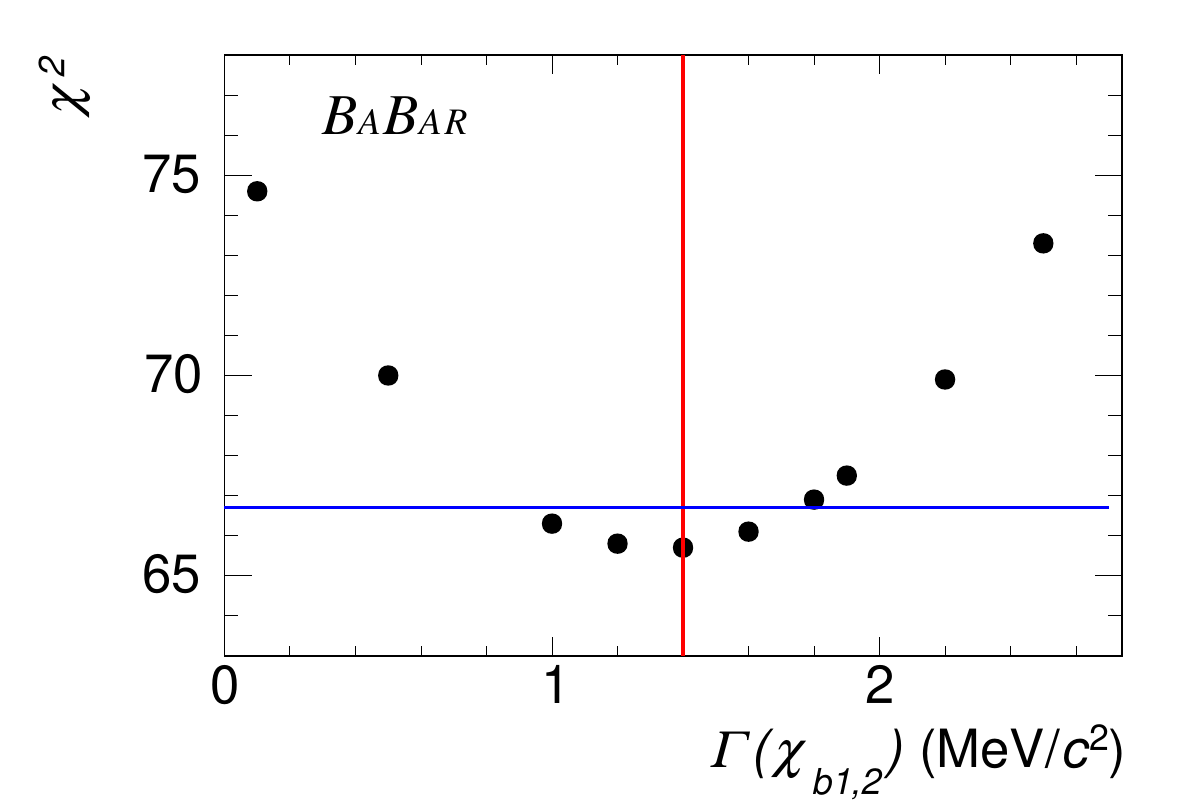}
\caption{Scan of the fit \chisq as function of the \itgamma  of the BW describing the \chibtot lineshapes. The horizontal line describes the change of \chisq by one unit.}
  \label{fig:fig10}
\end{center}
\end{figure}

Figure~\ref{fig:fig9}(a) shows the \mrec distribution for the total sample with the fitting function superimposed.
The fit yields $\chisq$/ndf=75/77, which statistically
 provides    
a good description of the data, but the visual comparison between the
data and the fitting function may indicate room for improvement. Therefore, the assumption of \mbox{\itgamma=0.1 \mev} in the BW function~Eq.(\ref{eq:bw}) is relaxed and a scan of the fit \chisq as a function of \itgamma is performed, with the result  
shown in Fig.~\ref{fig:fig10}. The distribution exhibits a well defined minimum
for \mbox{${\it \Gamma}=1.4 \pm 0.5$ \mev}. 

Similar disagreements between simulated and reconstructed $\chi_b$ lineshapes have been observed in the study of several radiative \TwoS and \ThreeS to $\chi_{bJ}(1P,2P)$ states~\cite{BaBar:2014och}. In this case the center-of-mass momentum of the $\gamma_s$ has been used and the broadening of the distributions with respect to the expected lineshapes from resolution functions was attributed to a Doppler effect due to the different rest frames involved in the decay chains.

The corresponding fit projections performed using this optimized \itgamma are shown in Fig.~\ref{fig:fig9}(b).
The \chibone and \chibtwo contributions, obtained using the weights $w(\chibone)$ and $w(\chibtwo)$ generated by the channel likelihood on the fit components are also shown.
Table~\ref{tab:tab2} gives the fractional contributions 
for this fit,  
which has a $\chisq$/ndf=66/77. The correlations between the different contributions are found to be $\chibtwo/\chibone = -0.90$ and $ps/\chibone,ps/\chibtwo=-0.22$.
\begin{table*} [!htb]
    \caption{Fitted fractions and events yields from the Channel Likelihood analysis of data, separately for the total, \mumu, and \ee samples with \itgamma=1.4 \mev.}  
    \label{tab:tab2}
    \begin{center}
\begin{tabular}{lrcrcrcr}
  \hline\\ [-2.3ex]
  Data  & Total & $c_1$ & \chibone  & $c_2$ & \chibtwo& $ps$ & $\chi^2$/ndf\cr
        & events  &      & events   &       &   events  &       &            \cr
  \hline\\ [-2.3ex]
Total & 1651 & $0.748 \pm 0.017$ &$1236 \pm 41$ & $0.223 \pm 0.017$ &$369\pm 29$& $0.028 \pm 0.007$ & 66/77  \cr
\mumu  & 1334 & $0.769 \pm 0.018$ &$1026 \pm 37$ & $0.203 \pm 0.018$ &$270 \pm 26$ & $0.028 \pm 0.008$ & 62/74\cr
\ee & 317 & $0.648 \pm 0.038$ & $205 \pm 17$ & $0.305 \pm 0.037$ & $97 \pm 13$ & $0.048 \pm 0.019$ & 23/31\cr
\hline
\end{tabular}
    \end{center}
\end{table*}

To quantitatively evaluate the possible presence of a \chibzero contribution, the dataset selection with no requirement on the presence of an $\omega$ signal is used, as discussed in Sec.~\ref{sec:event}. The description of the \chibzero is similar to that used for the \chibtot states but with a BW width fixed to 2.6 \mev\ as predicted in Ref.~\cite{Godfrey:2015dia}. The fit returns zero events for the \chibzero contribution with an uncertainty of 25 events for the total dataset and 22 events for the \mumu sample. These estimates are used to evaluate 90\% confidence level upper limits on the \chibzero yield of 41 and 36 events for the total and \mumu sample, respectively. The \chibzero upper limit contribution on the total dataset is indicated by the shaded region in Fig.~\ref{fig:fig6}(b).

\subsection{Measurement of the \boldmath{\chibtot} angular distributions}
\label{sec:ang}

\begin{figure}[!htb]
\begin{center}
  \includegraphics[width=0.49\textwidth]{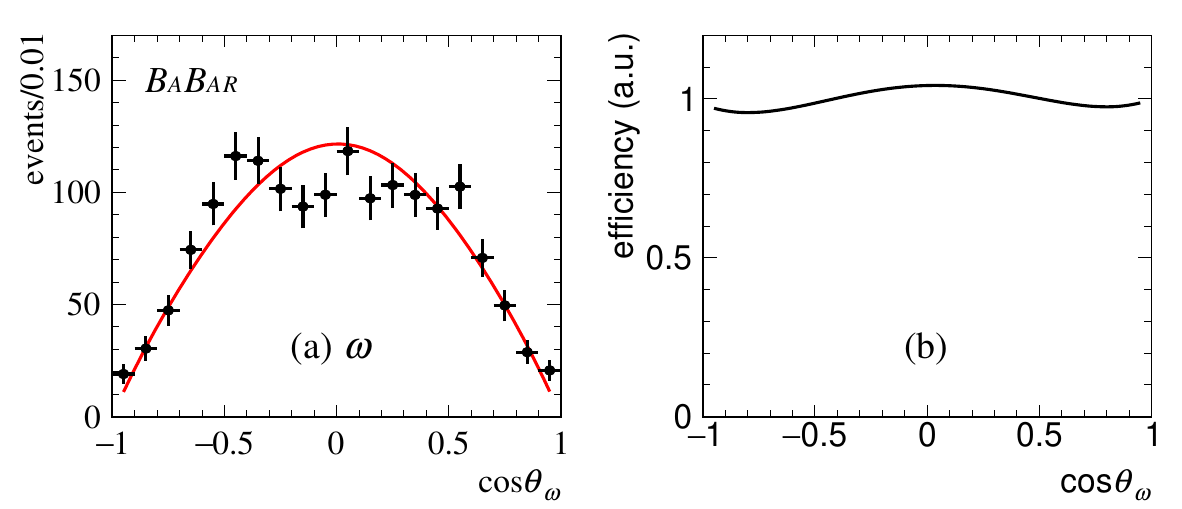}
  \includegraphics[width=0.49\textwidth]{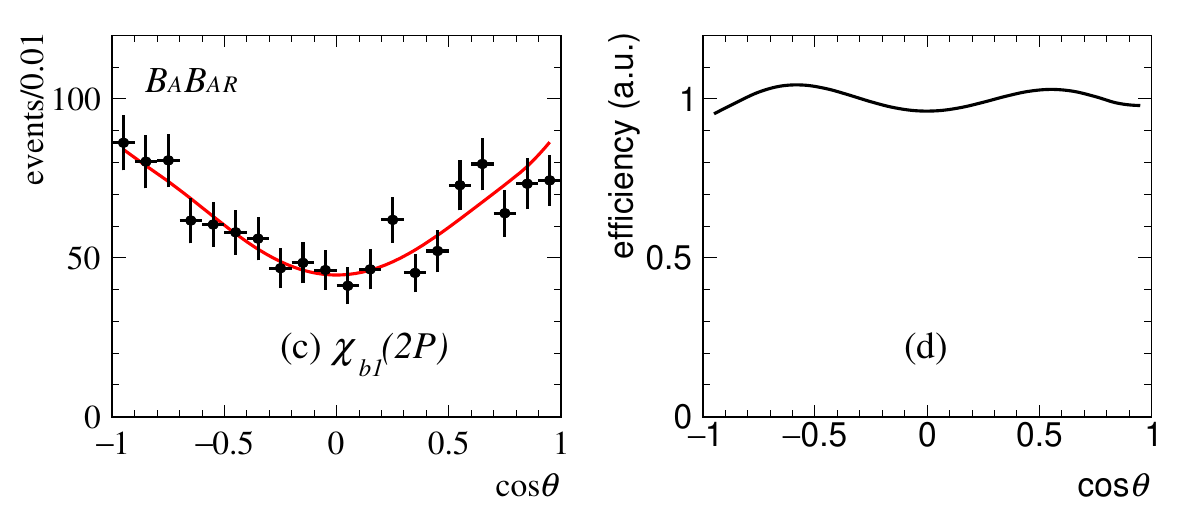}
    \includegraphics[width=0.49\textwidth]{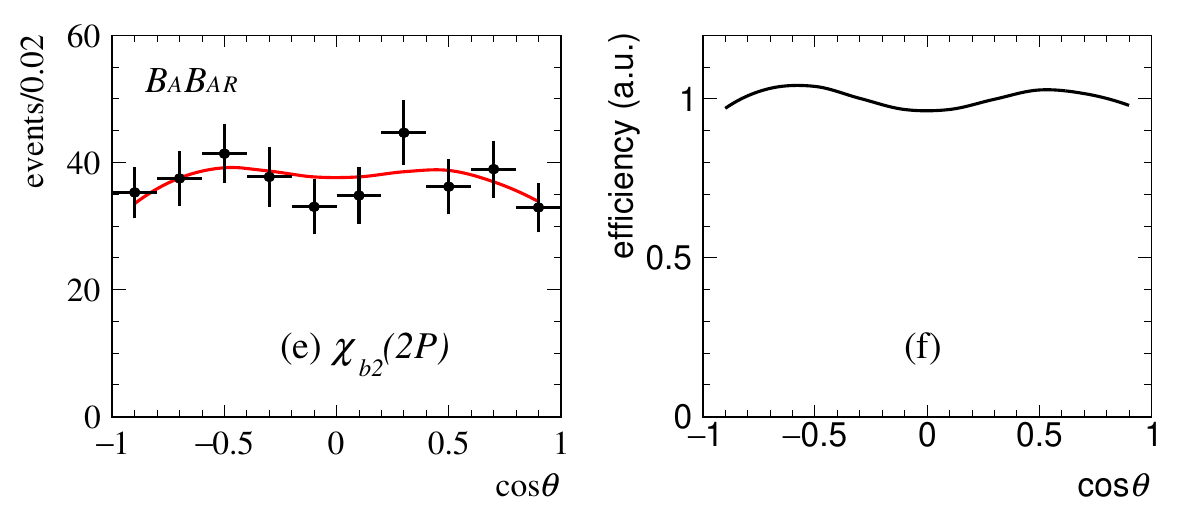}
    \caption{(a) Distribution of $\cos\theta_{\omega}$ for the sum of \chibone and \chibtwo weights. The curve is
      described by Eq.~(\ref{eq:womega}). (b) Corresponding efficiency distribution in arbitrary units (a.u.).
      (c) Distribution of $\cos\theta$ weighted by the \chibone weight. The curve is
      described by Eq.~(\ref{eq:wf1}). (d) Corresponding efficiency distribution. 
     (e) Distribution of $\cos\theta$ weighted by the \chibtwo weight. The curve is
      described by Eq.~(\ref{eq:wf2}). (f) Corresponding efficiency distribution.  
    }
  \label{fig:fig11}
\end{center}
\end{figure}

The fractional weights produced by the channel likelihood fit can be used to obtain information on the angular distributions describing the $\chibtot \to \omega \OneS$ decay.
Efficiency distributions as functions of $\cos \theta_{\omega}$ and $\cos \theta$ are evaluated separately for \mumu and \ee and for \chibone and \chibtwo MC samples. A smooth behavior is observed with little variations among the \chibone and \chibtwo mass regions, which are then combined also to reduce fluctuations due to the limited MC statistics. The resulting efficiency distributions are then smoothed by fitting forth order and sixth order polynomials as functions of $\cos\theta_{\omega}$ and $\cos \theta$, respectively. Angular efficiencies describing the total dataset are then obtained by averaging the \mumu and \ee efficiencies according to their data sizes.
The angular distribution for the $\omega \to \pip \pim \piz$ decay, a spin-one resonance, is expected to be described by~\cite{Lange:2001uf}
\begin{equation}
  W(\theta_{\omega}) =  1-\cos^2\theta_{\omega},
  \label{eq:w0}
\end{equation}
with the angle $\theta_{\omega}$ illustrated in Fig.~\ref{fig:fig7}(a). Figure~\ref{fig:fig11}(a)-(b) shows the distribution of $\cos \theta_{\omega}$, obtained
by weighting the events with the sum of the \chibone and \chibtwo weights $w(\chibone)+w(\chibtwo)$.

The superimposed curve is evaluated as
\begin{equation}
  f_{\omega} = W(\theta_{\omega})\cdot \eps(\theta_{\omega}),
  \label{eq:womega}
\end{equation}
where $\eps(\theta_{\omega})$ indicates the total efficiency function. 
An acceptable description of the data is obtained, with a resulting $\chisq$/ndf
of 42/19. The deviations of the data from the curve may be due
 to the presence of higher order physical effects, such as interfering intermediate $\rho \pi$ contributions~\cite{Nugent:2022ayu} not considered in the MC simulation.

The angular distribution for $\chibone \to \omega \OneS$ is expected to be described by~\cite{Voloshin:2003zz}
\begin{equation}
  W_1(\theta) = 1 + \cos^2 \theta, 
  \label{eq:w1}
\end{equation}
with the angle $\theta$ illustrated in Fig.~\ref{fig:fig7}(b).
Figure~\ref{fig:fig11}(c)-(d) shows the distribution of $\cos \theta$ obtained from weighting the events by the \chibone weight $w(\chibone)$.
The superimposed curve $f_{\chibone}$, given by Eq.~(\ref{eq:wf1}),  
is obtained by multiplying the expected functional form given by Eq.~(\ref{eq:w1}) by the efficiency $\eps(\theta)$. 
\begin{equation}
  f_{\chibone} = W_1(\theta)\cdot\eps(\theta).
  \label{eq:wf1}
\end{equation}
The model describes the data well with $\chisq$/ndf=15/19.  

The angular distribution for $\chibtwo \to \omega \OneS$ is expected to be described by~\cite{Voloshin:2003zz}
\begin{equation}
    W_2(\theta) = 1 - \frac{1}{7}\cos^2 \theta.
  \label{eq:w2}
\end{equation}
The \chibtwo signal is small and partially overlapping with the \chibone signal. To better separate the two channels, the description of the \chibone
is modified as
\begin{equation}
  f(\chibone) = BW_c(x)\cdot W_1(\theta)\cdot \eps(\theta). 
  \end{equation}
Figure~\ref{fig:fig11}(e)-(f) shows the distribution of $\cos \theta$ obtained from weighting the events by the \chibtwo weight $w(\chibtwo)$. 
The superimposed curve is found by multiplying the expected functional form~Eq.(\ref{eq:w2}) by the efficiency $\eps(\theta)$
\begin{equation}
  f_{\chibtwo} = W_2(\theta)\cdot\eps(\theta).
  \label{eq:wf2}
\end{equation}
Within the limited yield, the model describes the data well with $\chisq$/ndf=4/9.

As further test, the $\cos \theta$ efficiency corrected distributions are evaluated by dividing the measured distributions shown in Fig.~\ref{fig:fig11}(c) (Fig.~\ref{fig:fig11}(e)) by the corresponding efficiency shown in Fig.~\ref{fig:fig11}(d) (Fig.~\ref{fig:fig11}(f)) for \chibone and \chibtwo, respectively. The distributions are fitted using the function
  $W(\theta) = 1 + \alpha_J \cos^2 \theta$ obtaining values of  $\alpha_1=0.85\pm0.16$ and $\alpha_2=-0.12\pm0.16$ for \chibone and \chibtwo, respectively, in good agreement with the expected values of $\alpha_1=1$ and $\alpha_2=-0.14$.

\begin{figure}[!htb]
\begin{center}
  \includegraphics[width=0.49\textwidth]{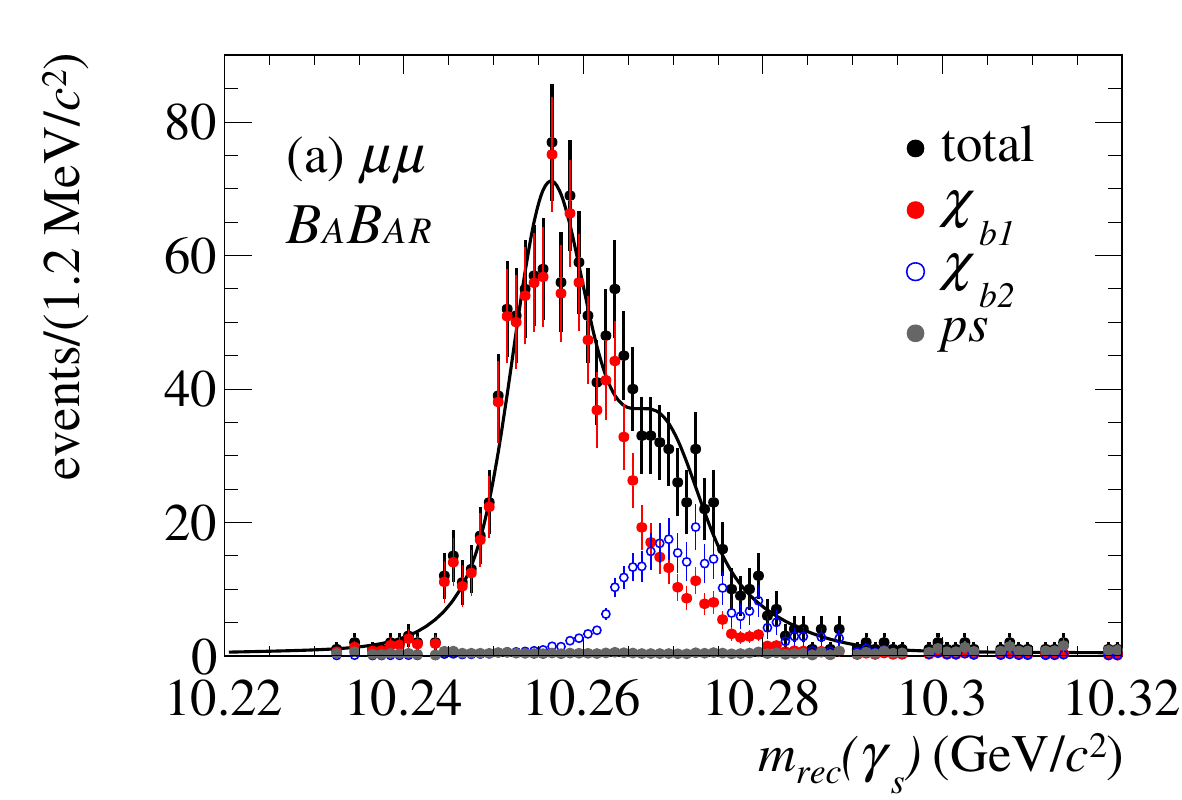}
  \includegraphics[width=0.49\textwidth]{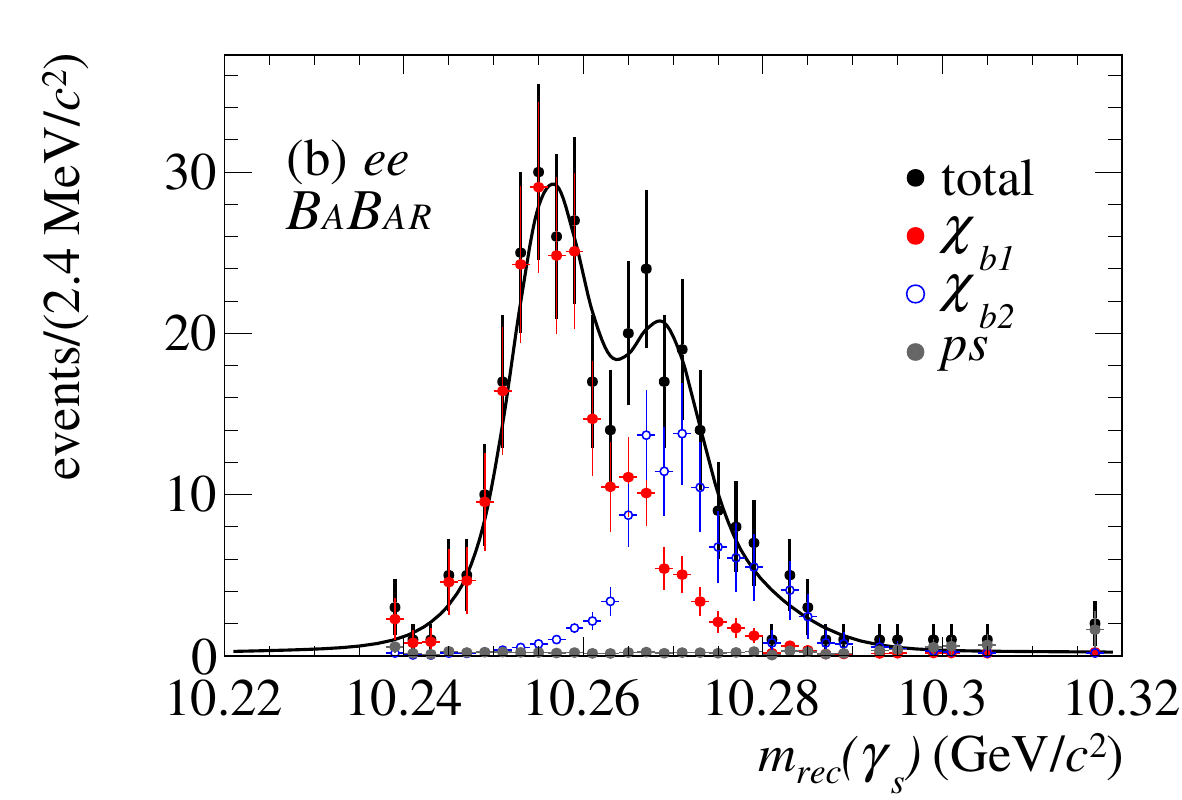}
\caption{Fit projections of the \mrec distribution for separate (a) \mumu and (b) \ee data. The total fitting function and the \chibtot components are superimposed.}
  \label{fig:fig12}
\end{center}
\end{figure}

Table~\ref{tab:tab2} presents details of the fit results for the total sample and separately for \mumu and \ee.
In these cases, separate efficiency distributions are used.
Figure~\ref{fig:fig12} shows the fit projections on the corresponding \mrec distributions.

\subsection{Measurement of the \boldmath{\chibtot} branching fractions}
\label{sec:br}

As discussed in Sec.~\ref{sec:effy}, only the \mumu data are used for the measurement of the \chibj branching fractions.
Table~\ref{tab:tab3} summarizes the information on the event yield and efficiency corrections needed to evaluate them.

\begin{table} [!htb]
  \caption{Summary of the information on the event yield and efficiency corrections used to evaluate the \chibtot branching fractions in the \mumu data.}
    \label{tab:tab3}  
    \begin{center}
\begin{tabular}{lcc}
  \hline\\ [-2.3ex]
      & \chibone & \chibtwo \cr
  \hline\\ [-2.3ex]
  Efficiency (\%)& $11.84\pm  0.09$ & $11.21 \pm  0.12$ \cr
  Event yield & $1026 \pm 37_{\rm stat} \pm 69_{\rm sys} $ & $270 \pm 26_{\rm stat} \pm 23_{\rm sys}$ \cr
  Corrected &  $8668 \pm 320_{\rm stat} \pm 581_{\rm sys}$ & $2408 \pm 233_{\rm stat} \pm 209_{\rm sys}$   \cr
  \hline
\end{tabular}
    \end{center}
\end{table}  

\subsection{Systematic uncertainties}
\label{sec:sys}

The list of systematic uncertainties is summarized in Table~\ref{tab:tab4}. Contributions are added in quadrature.
\begin{table} [!htb]
    \caption{Systematic uncertainties in the evaluation of the of the \chibtot fractions.}
    \label{tab:tab4}  
    \begin{center}
\begin{tabular}{lrr}
\hline\\ [-2.3ex]
Source & \chibone (\%) & \chibtwo (\%)\cr
\hline\\ [-2.3ex]
(a) Tracking efficiency& 2.5 & 2.5\cr
(b) $\gamma$ reconstruction efficiency & 1.8 & 1.8 \cr
(c) \piz\ reconstruction efficiency& 3.0 & 3.0 \cr
(d) Particle identification & 1.0 & 1.0 \cr
(e) Additional $\gamma$ and \piz\ & 0.6 & 0.7\cr
(f) Signal efficiency & 1.9 & 2.9 \cr
(g) \chibtot lineshape & 3.2 & 5.2 \cr
(h) \chibtot width & 2.9 & 2.6 \cr
(i) \TwoS background & 2.5 & 3.0 \cr
(j) $\omega$ selection & 1.5 & 3.1 \cr
\hline\\ [-2.3ex]
Total & 7.1 & 9.0 \cr
\hline      
\end{tabular}
\end{center}1
\end{table}

(a) Tracking efficiency. From detector studies based on control samples with high statistical precision, the tracking efficiency
for the decay $\ThreeS \to \pip \pim \OneS$, which has a similar kinematics, is found to be 2.5\%~\cite{BaBar:2011god}.

  (b)-(c) $\gamma$  and \piz\ reconstruction efficiencies. Similarly, a systematic uncertainty of 1.8 and of 3\% is assigned to the $\gamma$ and \piz\ reconstruction efficiency, respectively.

(d) Particle identification. Although particle identification is not used in this analysis, a 1\% uncertainty is
conservatively assigned on the separation between \mumu and \ee events.

(e) Additional $\gamma$ and \piz. The number of $\gamma_s$ candidates is increased to be $<9$ and the number of \piz\ candidates to $<7$. The differences with respect
to the reference selection are taken as systematic uncertainties.

(f) Signal efficiency. Separate efficiencies are evaluated for the \chibone and \chibtwo, listed in Table~\ref{tab:tab1} and found to differ by $4.2\sigma$.
 New estimates of the
weighted yields are obtained using the average efficiency.
The difference between the values obtained from the two methods is assigned as a systematic uncertainty.
An inspection of Fig.~\ref{fig:fig8} shows little dependence of the efficiency on angular variables. A test is made in which
the efficiency as a function of $\cos \theta$, fitted with a $6^{th}$ order polynomial, is used to weight the events.
This test yields only a 0.2\% difference with respect to the average efficiency,
  which is negligible.

  (g) \chibtot lineshape. An alternative model is used to describe the \chibtot lineshapes.
The MC truth-matched distributions of $m_{rec}(\gamma_s)$, obtained after all selection criteria are applied, are derived separately for the \chibone and \chibtwo and smoothed by fitting three Gaussian functions. The yield differences between the nominal and alternative model are taken as systematic uncertainties. In this case the comparison is made with respect to the non-optimized fitted values.

(h) \chibtot width. The differences between the fitted yields obtained assuming \itgamma=0.1 \mev\ and the optimized value of \itgamma=1.4 \mev\ are included as systematic uncertainties.

(i) \TwoS background. The antiselection on the background channels involving the \TwoS is based on the selection performed on $m_{rec}(\pip \pim)$. The simulation evaluates the loss of signal to be 12.1\%, 15.1\%, and 18.9\% for $2\sigma$, $2.5\sigma$ and $3.0\sigma$, respectively. The systematic uncertainty is evaluated varying the selection from $\pm 2\sigma$ to $\pm 2.5 \sigma$.

(j) $\omega$ selection. The low-mass tail of the $\omega$ is sensitive to the possible presence of a \chibzero signal.
The simulation evaluates the loss of the $\omega$ signal to be 0.5\%, 2.7\%, and 5.7\% for lower limits on the $\pip \pim \piz$ mass of 0.75, 0.76, and 0.765 \gevcc. The systematic uncertainty is evaluated varying the selection from 0.75 to 0.76 \gevcc.

Finally, in the evaluation of the branching fractions, the uncertainty in the number of collected \ThreeS events (see Eq.~(\ref{eq:bcount})) is included.
Additional tests performed by considering events having only one residual combination, or more than two combinations, are found to result in good agreement between the corresponding efficiency-corrected yields.

\subsection{Branching fractions}

The PDG~\cite{PDG} measurements used in this section are summarized in Table~\ref{tab:tab5}.
In the following, the ``pdg'' subscript is used to indicate uncertainties in known values.

\begin{table} [!htb]
    \caption{PDG branching fractions used in the present measurements. The $f_{\rm pdg}$ factor is defined by Eq.~(\ref{eq:br3}).}
    \label{tab:tab5}  
\begin{center}
\begin{tabular}{lcc}
\hline\\ [-2.3ex]
Decay  & \calB (\%) & $f_{\rm pdg}\times 10^{-3}$\cr
\hline\\ [-2.3ex]
$\ThreeS \to \gamma  \chibzero$ & \al $5.9\pm 0.6$ & $1.305 \pm 0.135$\cr
$\ThreeS \to \gamma  \chibone$ & $12.6\pm 1.2$ & $2.787 \pm 0.270$\cr
$\ThreeS \to \gamma  \chibtwo$ & $13.1\pm 1.6$ & $2.898 \pm 0.360$\cr
$\OneS \to \mup \mun$ & \al$2.48\pm 0.05$ \cr
$\omega \to \pip \pim \piz$ & $89.2\pm0.7$ \cr
\hline
\end{tabular}
\end{center}
\end{table}
The $\chibj \to \omega \OneS$ branching fractions are evaluated as
\begin{equation}
  \calB(\chibj\to\omega\OneS) = \frac{N_{\rm corr}(\chibj\to\omega\OneS)}{N(\chibj)}
      \label{eq:br1}
\end{equation}
in the decays $\ThreeS \to \gamma \chibj$.
Here, for each $\chibj$, $N_{\rm corr}(\chibj\to\omega\OneS)$ is the number of measured events corrected for efficiency and unseen decay modes and $N(\chibj)$ is the total $\chib$ yield produced in \ThreeS\ decays,
\begin{equation}
  N(\chibj) = N_{\ThreeS} \cdot \calB(\ThreeS \to \gamma \chibj),
  \label{eq:br2}
\end{equation}
where $N_{\ThreeS}$ is the total available \ThreeS\ yield evaluated using the method of $B$ counting~\cite{BaBar:2014omp}:
 \begin{equation}
    N_{\ThreeS} = (121.3 \pm 1.2_{\rm sys}) \times 10^6.
    \label{eq:bcount}
 \end{equation}
 Grouping the PDG factors, Eq.~(\ref{eq:br1}) can be written as
\begin{equation}
  \calB(\chibj\to\omega\OneS) = \frac{N(\chibj\to\omega\OneS)}{\eps(\chibj)\cdot N_{\ThreeS} \cdot f_{\rm pdg}},
      \label{eq:br3}
\end{equation}
where, for each $\chibj$, $N(\chibj\to\omega\OneS)$ is the number of measured events, $\eps(\chibj)$ is the efficiency, and $f_{\rm pdg}$ are the PDG correction factors
\begin{equation}
  \begin{split}
    &  f_{\rm pdg} = \\
     & \calB(\omega\to\pip \pim \piz)\cdot\calB(\OneS \to \mup \mun)\cdot\calB(\ThreeS \to \gamma \chibj)
   \end{split}   
\end{equation}
listed in Table~\ref{tab:tab5}.

  Using Eq.~(\ref{eq:br3}), the branching fractions are determined to be
  \begin{equation}
    \begin{split}
      & \calB(\chibone\to\omega\OneS) = \\
      & (2.56 \pm  0.09_{\rm stat} \pm 0.18_{\rm sys} \pm 0.25_{\rm pdg})\%
       \end{split}
       \label{eq:chib1}
   \end{equation}
  and
  \begin{equation}
    \begin{split}
       & \calB(\chibtwo\to\omega\OneS) =\\        
       & (0.69 \pm  0.07_{\rm stat} \pm 0.06_{\rm sys} \pm 0.09_{\rm pdg})\%.
       \end{split} 
    \label{eq:chib2} 
  \end{equation}
    The ratio $r_{2/1}$ is also evaluated, with most of the systematic uncertainties canceling out, except for items (f), (g), and (h) in Table~\ref{tab:tab5}
    \begin{equation}
      \begin{split}
        r_{2/1} = & \frac{\calB(\chibtwo \to \omega \OneS)}{\calB(\chibone \to \omega \OneS)} \\
        = & 0.27 \pm 0.03_{\rm stat} \pm 0.02_{\rm sys}\pm 0.04_{\rm pdg}.
        \end{split}
     \label{eq:r12} 
    \end{equation}
    These measurements are in agreement with the results from the CLEO~\cite{CLEO:2003vio} and Belle~\cite{Belle:2024azd} Collaborations but with significantly better precision.

No evidence is found for a $\chibzero\to\omega\OneS$ decay. 
To measure the \chibzero branching fraction upper limit for this decay mode,
the estimated limit on the \chibzero yield of 36 events discussed in Sec.~\ref{sec:chan} is inserted in Eq.~(\ref{eq:br3}) using the average \chibtot efficiency listed in Table~\ref{tab:tab1} to obtain 
\begin{equation}
  \calB(\chibzero\to\omega\OneS) < 0.23\% \ \rm{at} \ 90\% \ {\rm C.L.} 
  \end{equation}
    
    \section{Summary}

Results are presented on $\chi_{b1,2}(2P) \to \omega \Upsilon(1S)$ transitions from $\epem\to\Upsilon(3S) \to \gamma \chi_{b1,2}(2P)$ decays. The data were collected
with the \babar\ detector at the PEP-II asymmetric-energy collider. The integrated
luminosity of the sample is 28.0 fb$^{-1}$, corresponding to $121.3 \times 10^6$ $\Upsilon(3S)$ decays.    
Clean $\chi_{b1,2}(2P)$ signals are observed and improved precision measurements of branching fractions are derived as:
\begin{equation}
   \begin{split}
    & \calB(\chibone\to\omega\OneS) =\\
    & (2.56 \pm  0.09_{\rm stat} \pm 0.18_{\rm sys} \pm 0.25_{\rm pdg})\%
     \end{split}
\nonumber
  \end{equation}
  and
  \begin{equation}
     \begin{split}
       &  \calB(\chibtwo\to\omega\OneS) = \\
       & (0.69 \pm  0.07_{\rm stat} \pm 0.06_{\rm sys} \pm 0.09_{\rm pdg})\%.
        \end{split}
      \nonumber
    \end{equation}
    The ratio $r_{2/1}$ is also evaluated
    \begin{equation}
        r_{2/1} = 0.27 \pm 0.03_{\rm stat} \pm 0.02_{\rm sys}\pm 0.04_{\rm pdg}.
\nonumber
    \end{equation}
    Theoretical expectations on this ratio are given in Ref.~\cite{Voloshin:2003zz}, which states
    that the expected ratio of the absolute decay rates of
    $\Gamma(\chibtwo \to \omega \OneS)/\Gamma(\chibone \to \omega \OneS)$ is  given by the ratio of the S-wave phase-space factors, which is approximately 1.4. The predicted value for this ratio is 
    $r_{2/1} \approx 1.3 \pm 0.3$ where the quoted uncertainty arises mainly from the knowledge of 
    total decay rates of the $\chi_{bJ}$ resonances. Adding statistical and systematic uncertainties in the present measurement,
    a difference of $\approx 3.4 \sigma$ is found with respect to the predicted value.
   The results are consistent with previous measurements from CLEO~\cite{CLEO:2003vio} and Belle~\cite{Belle:2024azd}, with significantly better precision.

    Angular distributions for $\chibone \to \omega \OneS$ and $\chibtwo \to \omega \OneS$ are measured for the first time and are found to be 
    in agreement with theoretical expectations~\cite{Voloshin:2003zz}.
No evidence is found for the presence of a $\chibzero \to \omega \OneS$ decay mode, and an upper bound of 
$\calB(\chibzero\to\omega\OneS) < 0.23\% \ {\rm at} \ 90\% \ {\rm C.L.}$  is set.\\

\appendix

\section{Study of the \boldmath{\ThreeS \to \pip \pim \OneS} decay}
\label{sec:appendix}

The validation of the performance of the MC simulation of the background $\gamma$'s from final state radiation effects is performed by comparing data and simulation on a reference \ThreeS decay mode which does not involve the presence of $\gamma$'s in the decay.
The decay 
\begin{equation}
\ThreeS \to \pip \pim \OneS
  \label{eq:ref}
\end{equation}
where $\OneS \to \mup \mun$ is reconstructed from data and simulation.
For the data, only events with tagged \mumu candidates are selected while, for simulation, MC generated and reconstructed events  for decay~(\ref{eq:ref}) are used.

\begin{figure*}[!htb]
\begin{center}
  \includegraphics[width=0.40\textwidth]{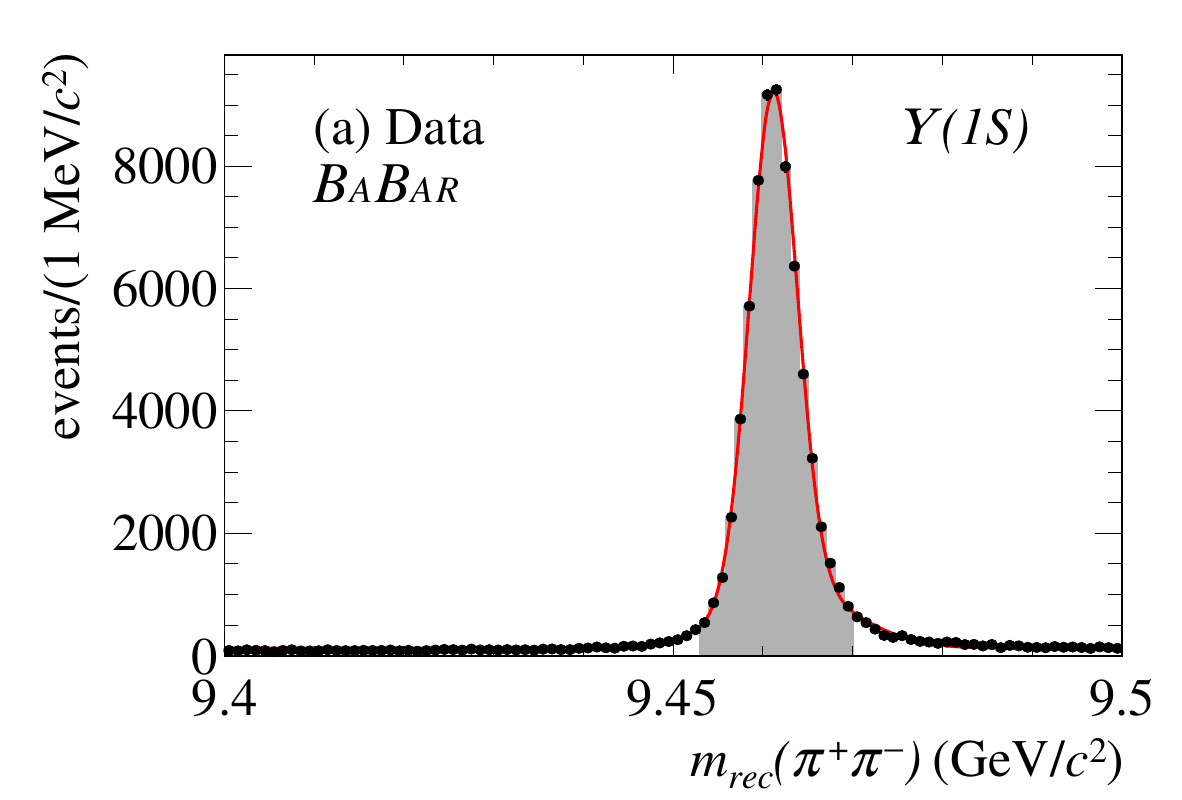}
  \includegraphics[width=0.40\textwidth]{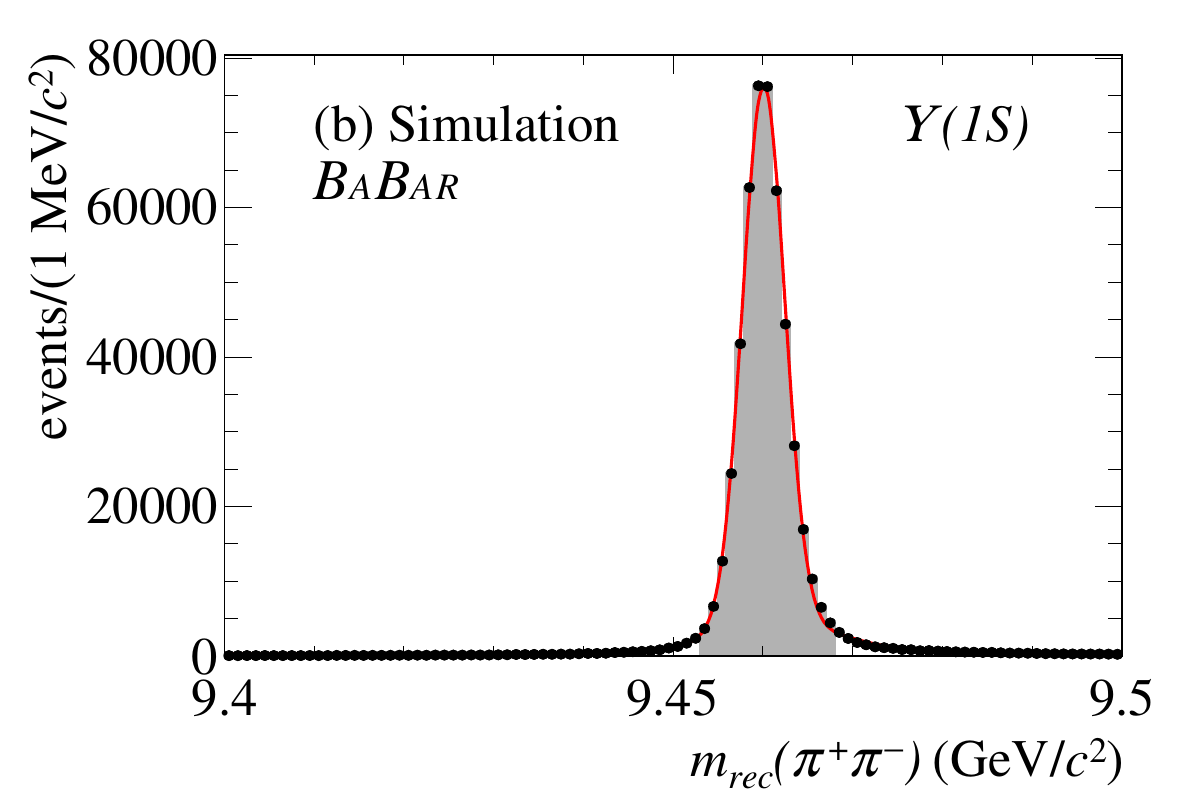}
  \caption{Recoil mass to the $\pip \pim$ system $m_{rec}(\pip \pim)$ for: (a) data and (b) MC simulation, for the decay channel~(\ref{eq:ref}). The shaded area shows the region of the selected \OneS candidates.}
\label{fig:fig13}
\end{center}
\end{figure*}

Similarly to the procedure described in Sec.~\ref{sec:reco}, the two fastest tracks with momentum $p>2.8$ GeV/c are assumed to be muons while the slowest tracks with momentum $p<1.1$ GeV/c are assumed to be pions.
Figure~\ref{fig:fig13} shows the recoil mass $m_{rec}(\pip \pim)$ (Eq.(\ref{eq:mrec_pipi})) for data and signal MC simulation, where clean \OneS signals can be observed. The \OneS signal is then selected in the $\pm 3\sigma_e$ mass region of the \OneS mass, where \mbox{$\sigma_e=3.53 \pm 0.17$ \mevcc} and \mbox{$\sigma_e=2.78 \pm 0.03$ \mevcc} for data and MC simulation, respectively.
The larger \OneS width in the data can be understood as due to the presence of several final states contributions
in the inclusive \OneS production.
Momentum balance is applied similar to that used in Eq.~(\ref{eq:chi1}) but considering only the four charged tracks from the \ThreeS decay.
Events within $\pm 3 \sigma$ are selected, where $\sigma_{xy}=37$ \mevc\ and \mbox{$\sigma_z=52$ \mevc.}

\begin{figure*}[!htb]
\begin{center}
  \includegraphics[width=0.40\textwidth]{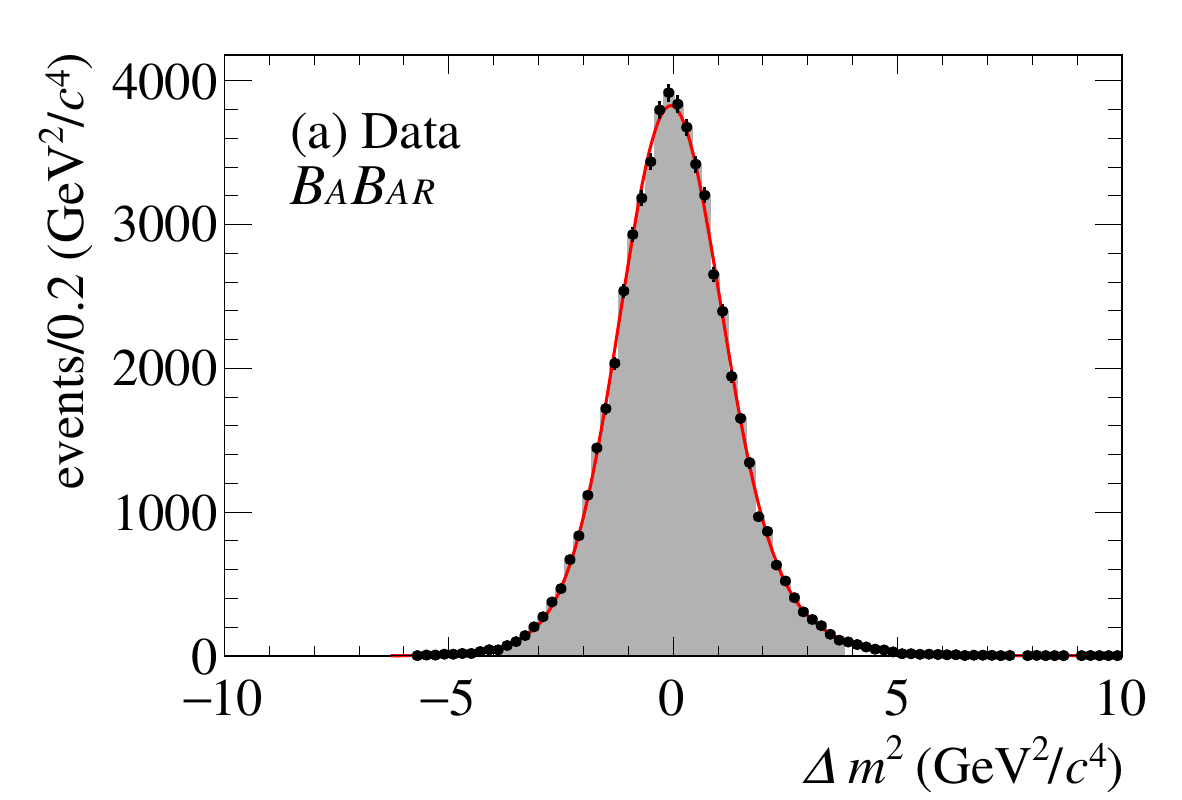}
  \includegraphics[width=0.40\textwidth]{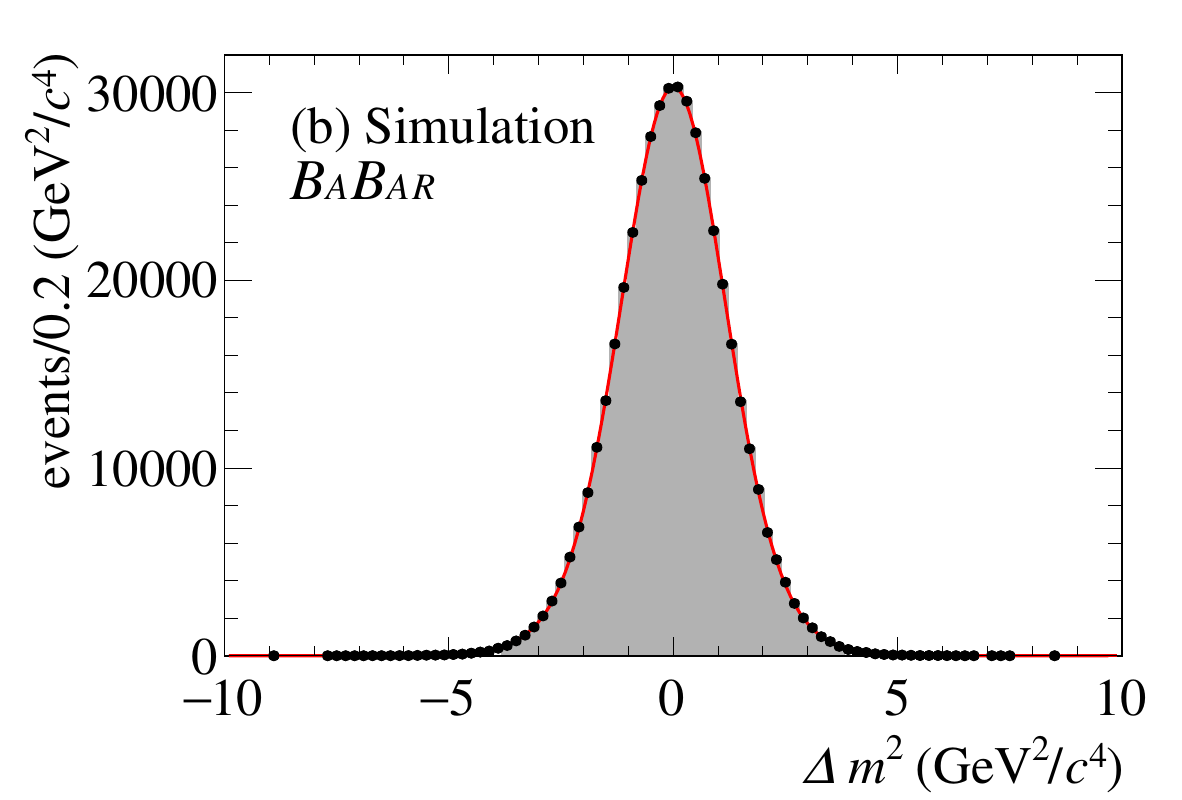}
  \caption{Distributions of $\itdelta m^2$ for: (a) data and (b) MC simulation, for the decay channel~(\ref{eq:ref}). The shaded area shows the region of the selected $\ThreeS \to \pip \pim \OneS$ candidates.}
\label{fig:fig14}
\end{center}
\end{figure*}

The final selection is performed on the variable
\begin{equation}
  \itdelta m^2 = m^2_{rec}(\pip \pim) - m^2(\mumu) 
\end{equation}
shown in Fig.~\ref{fig:fig14} for data and MC. A fit to the distributions using the sum of two Gaussian functions sharing the same mean gives $\sigma_e=1.28 \pm 0.05$ GeV$^2/c^4$ and $\sigma_e=1.25 \pm 0.05$ GeV$^2/c^4$ for data and MC simulation, respectively. As consistent resolutions are observed in $\ThreeS \to \pip \pim \OneS$, this analysis
validates the agreement between data and MC simulations. 
Selecting events within $\pm 3\sigma_e$ total yields of \mbox{$(58.4 \pm 0.2)\times 10^3$} and $(461.7 \pm 0.7)\times 10^3$ are obtained for data and MC simulation, respectively.

\begin{figure*}[!htb]
\begin{center}
  \includegraphics[width=0.90\textwidth]{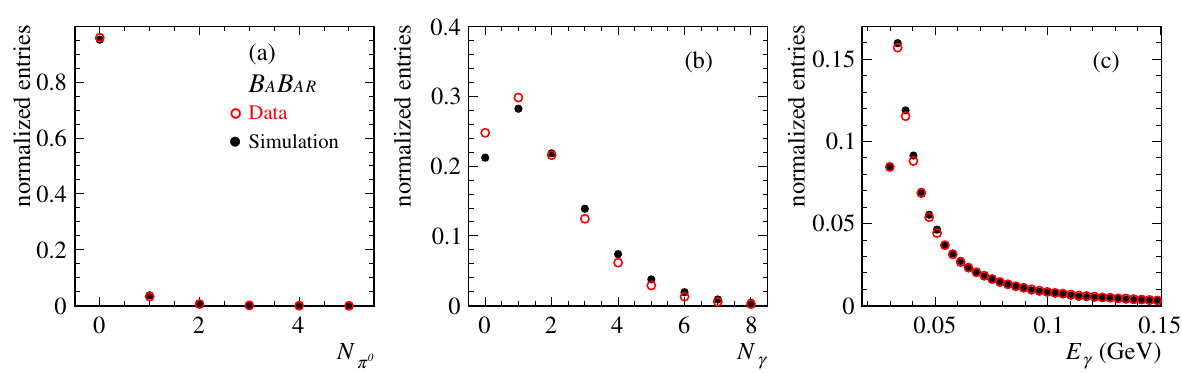}
  \caption{Normalized distributions of (a) the number of candidate \piz's, (b) the number of candidate $\gamma$'s and (c) the $\gamma$'s energy distribution for data and simulation from the reconstructed $\ThreeS \to \pip \pim \OneS$ candidates.}
\label{fig:fig15}
\end{center}
\end{figure*}
For the selected events no $\gamma$'s are present in the reconstructed \ThreeS decay mode and therefore all the activity present in the EMC should be due to final state radiation or incident beams background.  Figure~\ref{fig:fig15} shows a comparison between the normalized distributions
of (a) the number of candidate \piz's, (b) the number of candidate $\gamma$'s and (c) the $\gamma$'s energy distribution for data and simulation. A good agreement is observed, validating the MC performance in the soft photons background simulation.

\clearpage

\section{Acknowledgments}

\input{babar_acknowledgements_reduced_mar2023.txt}

\renewcommand{\baselinestretch}{1}

\end{document}

%% file: authors_aug2025_frozen_no_institutes.tex
\author{J.~P.~Lees}
\author{V.~Poireau}
\author{V.~Tisserand}
\author{E.~Grauges}
\author{A.~Palano}
\author{G.~Eigen}
\author{D.~N.~Brown}
\author{Yu.~G.~Kolomensky}
\author{M.~Fritsch}
\author{H.~Koch}\thanks{Deceased}
% ...
\collaboration{The \babar\ Collaboration}
\noaffiliation

%% file: babar_acknowledgements_reduced_mar2023.txt
We are grateful for the extraordinary contributions of our PEP-II colleagues in achieving the excellent luminosity and machine conditions that have made this work possible. The success of this project also relies critically on the expertise and dedication of the computing organizations that support \babar, including GridKa, UVic HEP-RC, CC-IN2P3, and CERN. The collaborating institutions wish to thank SLAC for its support and the kind hospitality extended to them. We also wish to acknowledge the important contributions of J.~Dorfan and our deceased colleagues E.~Gabathuler, W.~Innes, D.W.G.S.~Leith, A.~Onuchin, G.~Piredda, and R. F.~Schwitters.